\documentclass[fleqn,usenatbib]{mnras}

\usepackage{bm}

\usepackage[T1]{fontenc}
\usepackage[utf8]{inputenc}

\DeclareRobustCommand{\VAN}[3]{#2}
\let\VANthebibliography\thebibliography
\def\thebibliography{\DeclareRobustCommand{\VAN}[3]{##3}\VANthebibliography}

\usepackage{graphicx}	% Including figure files
\usepackage{amsmath}	% Advanced maths commands
\usepackage{amssymb}	% Extra maths symbols
\usepackage[varvw]{newtxmath}  %Added to prevent cursive v to look like $\nu$.
\usepackage{ulem}

\newcommand{\aM}{$\alpha$-$M$}
\newcommand{\av}{\alpha_{\rm vir}}
\newcommand{\ceff}{c_{\rm eff}}
\newcommand{\cs}{c_{\rm s}}

\newcommand{\Dtsf} {\Delta t_{\rm sf}}
\newcommand{\Eg}{E_{\rm g}}
\newcommand{\Ek}{E_{\rm k}}
\newcommand{\epsff}{\epsilon_{\rm ff}}
\newcommand{\hii}{{\sc{Hii}}}
\newcommand{\kms}{\mbox{\,km s}^{-1}}
\newcommand{\Lcal} {{\cal L}}
\newcommand{\LS}{$\Lcal$-$\Sigma$}
\newcommand{\Mg}{M_{\rm g}}
\newcommand{\MJ}{M_{\rm J}}
\newcommand{\MJt}{M_{\rm J,t}}
\newcommand{\Ms}{M_{\rm s}}

\newcommand{\Msun}{\mbox{\,M}_\odot}
\newcommand{\pcc}{\mbox{\,cm}^{-3}}
\newcommand{\psc}{\mbox{\,cm}^{-2}}

\newcommand{\tff}{\tau_{\rm ff}}
\newcommand{\tffloc}{\tau_{\rm ff,loc}}

\newcommand*{\vcenteredhbox}[1]{\begin{tabular}{@{}c@{}}#1\end{tabular}}

\usepackage[dvipsnames]{xcolor}

\title[The TS and GHC models compared]{The Turbulent Support (TS) and Global Hierarchical Collapse (GHC) models for molecular clouds compared. Differences, convergence, and myths}

\author[V\'azquez-Semadeni et al.]{Enrique V\'azquez-Semadeni$^1$, Aina Palau$^1$, Gilberto C.\ G\'omez$^1$, \newauthor Griselda Arroyo-Ch\'avez$^1$, Christian Alig$^{2,3}$, Javier Ballesteros-Paredes$^1$, \newauthor 
Vianey Camacho$^{1,4}$, Alessio Traficante$^5$, Alejandro Gonz\'alez-Samaniego$^6$, \newauthor Manuel Zamora-Avil\'es$^7$, and Andreas Burkert$^{2,3}$
\\
\\
$^1$Instituto de Radioastronomía y Astrofísica,
      Universidad Nacional Aut\'onoma de M\'exico, Apdo. postal 3-72, Morelia Mich. 58089, M\'exico \\
      $^2$Universit\"ats-Sternwarte, Fakult\"at f\"ur Physik, Ludwig-Maximilians-Universit\"at M\"unchen, Scheinerstraße 1, D-81679 M\"unchen, Germany\\
      $^3$Excellence Cluster ORIGINS, Boltzmannstr. 2, D-85748, Garching, Germany\\
      $^4$Center of Astronomy and Gravitation, Department of Earth Sciences, National Taiwan Normal University, 88, Sec. 4, Ting-Chou Rd., Wenshan District, Taipei 116, Taiwan R.O.C\\
      $^5$INAF-Istituto di Astrofisica e Planetologia Spaziali (INAF-IAPS), Via Fosso del Cavaliere 100, I-00133, Roma, Italy\\
      $^6$Universidad Aut\'onoma Metropolitana\\
      $^7$Instituto Nacional de Astrof\'isica, \'Optica y Electr\'onica\\
}

\date{Accepted XXX. Received YYY; in original form ZZZ}

\pubyear{2022}

\begin{document}

\maketitle

\begin{abstract}
We provide a detailed comparison between the ``turbulent support'' (TS) and ``global hierarchical collapse'' (GHC) models for molecular clouds and star formation, their respective interpretations of the observational data, the features they share, and suggested tests and observations to discern between them. Also, we clarify common misconceptions in recent literature about the global and hierarchical nature of the GHC scenario, and briefly discuss the evolution of some aspects of both models toward convergence. TS assumes that star-forming molecular clouds and their substructures are either in approximate virial equilibrium between gravity and turbulence or overvirial, so that the cloud is either confined or expanding, and its substructures (clumps, filaments and cores) are produced by turbulent compressions. In this scheme, the star formation rate (SFR) is time-independent and determined by the turbulent and gravitational parameters of the clouds, in particular the virial parameter $\av$. Conversely, GHC assumes that most star-forming molecular clouds and their substructures are part of a continuous gravitationally-driven flow, each accreting from their parent structure. Therefore, GHC is an intrinsically {\it evolutionary} model for the clouds and their star formation rate, determined by the evolution of the collapse flow. It interprets nonthermal motions as a mixture of infall and turbulent components, with the relative importance of the former increasing as the objects become denser and/or more massive, and thus $\av$ is an {\it evolving variable} of the clouds. Tests that may provide clues to distinguishing between TS and GHC must take into account that the innermost parts of globally gravitationally bound structures may not locally appear bound, and thus the binding may have to be searched for at the largest scale of the structure.

%- TS does not take into account cloud evolution
%- TS consider equilibrium but this is not valid cloud by cloud, only statistically; circulation maybe only in the outflow feedback
%- linewidth is not fully turbulent

\end{abstract}

\section{Introduction}

Molecular clouds (MCs) are the densest regions of the interstellar medium (ISM), and constitute an ``anomalous'' phase of the ISM in the sense that they have much higher thermal pressures ($> 10^4$ K cm$^{-3}$) than the mean thermal Galactic midplane value \citep[$\sim 3 \times 10^3$ K cm$^{-3}$;] [] {Boulares_Cox90}, which the other phases share.
%{\color{magenta} JBP: Abogadodeldiablo: ahora que he estado revisando a Ostriker+10, medio argumentan que lo que pasa es que la presión turbulenta y la magnética tmbn contribuyen, de manera que las NMs estarían en equilibrio de presión total... }
Also, they are known to contain hundreds or thousands of thermal Jeans masses,\footnote{For example, at density $n = 300\, \pcc$ and temperature $T = 15$ K, the Jeans mass is $\MJ \sim 50 \Msun$, and so a $10^5$-$\Msun$ giant molecular cloud contains $\sim 2000$ Jeans masses.} and so they must be either collapsing gravitationally, or supported by some mechanism against collapse, such as magnetic or turbulent support, but nevertheless in a highly compressed state in response to gravity. MCs are also known to contain strongly supersonic motions, implying sonic Mach numbers $\Ms \sim 5$--20 \citep[e.g.,] [] {Wilson+70, Larson81, Solomon+87}.
%The notion of MCs undergoing collapse was dismissed several decades ago \citep{Zuck_Palmer74} under the argument that it would produce much larger star formation rates (SFRs) than observed, although of course another possibility is that, if molecular clouds are undergoing global collapse, they may be destroyed by the massive stars formed within them long before most of their mass is turned into stars \citep[e.g.,] [] {Field70}.

In the last couple of decades, 
%the possibility that MCs are magnetically supported has been mostly abandoned, since it has been observationally found that most molecular clouds and their substructures have supercritical mass-to-magnetic flux ratios \citep{Crutcher12, Hu+23, Pattle+23}, and so 
two main models have become predominant for the description of MC structure, dynamics, and their star formation activity. On the one hand, there is the {\it turbulent support} (TS) model, which posits that MCs are nearly hydrostatic structures, with lifetimes much longer than their global free-fall time, supported against their self-gravity by virialized supersonic turbulence (perhaps assisted by magnetic fields)\citep[e.g.,] [] {Zuck_Evans74, Larson81, VS+00, MacL_Kles04, BP+07, McKee_Ostr07, Henneb_Falg12, Kles_Glov15} and the {\it global hierarchical collapse} (GHC) model, which posits that MCs and their substructures are in general undergoing continuous accretion from their environment and an infall flow driven by self-gravity at their star-forming stages, and that the main effect of stellar photoionizing feedback is to disrupt the clouds long before most of their mass can be turned into stars, rather than maintaining them in near equilibrium \citep[e.g.,] [] {Hoyle53, Field70, Liszt+74, Goldreich_Kwan74, HBB01, Hart_Burk07, Heitsch+08a, Heitsch+08b, Heitsch_Hart08, VS+09, VS+19, BP+11, IbanezM+16, IbanezM+17, MacLow+17}. Several observational works are now aimed at discerning between the two. 

Within the context of the assumption of supported clouds in the TS scenario, and the resulting clump mass function (CMF), several theoretical works have proposed that the star formation rate (SFR) and the stellar initial mass function (IMF) are essentially regulated by turbulent parameters such as the {\it sonic and Alfv\'enic Mach numbers}, the {\it virial parameter} (eq.\ \ref{eq:Larson_vir} below),  the compressibility parameter, and the equilibrium pressure in the midplane of disks galaxies \citep[e.g.,] [] {Pad_Nord02, Krum_McKee05, Henneb_Chab08, Henneb_Chab09, Ostriker+10, Henneb_Chab11, Henneb_Chab13, Pad_Nord11, Federrath_Klessen12}.
%and the origin of the stellar initial mass function (IMF) as a consequence of the clump mass function \citep[CMF; e.g.,] [] {Pad_Nord02, Henneb_Chab08, Henneb_Chab09, Henneb_Chab13, Hopkins12}. %
Also, on the basis of the TS scenario, \citet{McKee_Tan02} and \citet{McKee_Tan03} generated a model in which the high accretion rates onto protostellar objects within massive clumps resulted from the large turbulent pressure generated by the strongly supersonic motions within these clumps. Furthermore, these strongly supersonic motions, interpreted as turbulence, are also at the basis of the notion of a {\it turbulent Jeans mass} \citep{MacL_Kles04} as the fragmentation criterion in massive cores. 

On the other hand, under the scope of the GHC scenario, various models and studies have been presented, including: the general scenario for molecular cloud formation and subsequent contraction \citep{HBB01}; a model for the simultaneous {\it evolution} of the SFR and the clouds' masses \citep{ZA+12, ZA_VS14}; a model for the evolution of the density profiles of collapsing cores \citep{Gomez+21} and for the simultaneous evolution of the cores' masses and the mass of the stars needed to disrupt the accretion rate onto them \citep{VS+24}; a study on the dependence of the star formation efficiency on the gas surface density as a consequence of the shorter contracton timescales for denser gas \citep{Burk_Hart13}; a model for the upper end of the IMF based on a distribution of accretion rates arising from gravitationally focused accretion \citep{BP+15}; analytical \citep{Heitsch13} and numerical \citep{GV14} models for the gravitational infall onto molecular cloud filaments; and a comparison on the cloud formation and destruction timescales \citep{MacLow+17}.

\subsection{Historical perspective and context} \label{sec:history}

Early interpretations of the observed supersonic linewidths in MCs were that the clouds were probably undergoing gravitational contraction, or at least some form of cloud-scale compression \citep[or expansion;] [] {Liszt+74, Goldreich_Kwan74}. However, this suggestion was soon dismissed \citep{Zuck_Palmer74, Zuck_Evans74} under the arguments that {\it a)} it would imply too large a star formation rate; {\it b)} it would imply the existence of systematic shifts between emission lines produced at the clouds' centers and absorption lines produced at the clouds' edges, which were however not observed, and {\it c)} it was not clear what the center of the collapse could be, as clouds contain multiple dense condensations, clusters, and \hii\ regions. In view of these considerations, \citet{Zuck_Evans74} suggested that the amplitude of the cloud-scale velocity differences (or simply, of the cloud-scale {\it motions}) are generally small in comparison with the amplitude of the small-scale, turbulent ones. This implied that the turbulent motions had to have size scales much smaller than the clouds themselves. Thus started the scenario of globally supported MCs, with the main supporting agents being either the magnetic field \citep[e.g.,] [] {Shu+87, Mouschovias91} or MHD supersonic turbulence \citep[e.g.,] [] {Pad_nord99, VS+00, VS+03, MacL_Kles04, Ballesteros-Paredes+07, McKee_Ostr07, Henneb_Falg12}. 

However, over the last one and a half decades, advances in our understanding of turbulence and of the nature of gravitationally-driven motions have prompted a return to considering cloud-scale contraction as a viable possibility for the dynamical state of MCs. These have been discussed in \citet{VS+19}, and for the most part we do not repeat them here, although we do wish to emphasize that gravitational contraction provides a natural mechanism for generating near equipartition between the gravitational and kinetic energy at all scales, while there is no known mechanism that would allow turbulence to approach virialization at all scales within molecular clouds. This is because fluid turbulence is a strongly dissipative process, and needs constant replenishment from external sources to be maintained at just the precise rate at every scale to maintain virialization. However, any sources of kinetic energy other than gravitational contraction itself are unrelated to the gravitational energy of the clouds, and therefore can greatly exceed, or be exceeded by, it. 

Indeed, numerical simulations including feedback show that it can either be insufficient to stop gravitational contraction at the cloud scale \citep[see, e.g., the review by] [] {Dale15}, or be capable of completely destroying the clouds \citep[e.g.,] [] {Colin+13, Kim+21}, depending on the feedback mechanisms and cloud shapes and masses considered. Also, recent observational determinations of the life cycle of clouds suggest that pre-supernova feedback (mostly from photoionising UV radiation) generated {\it within} the clouds destroys the clouds \citep[e.g.,] [see also \citealt{MacLow+17} for numerical evidence to the same] {Chevance+22} rather than maintain them in near equilibrium. Conversely, supernova feedback generated {\it outside} the clouds is more likely to {\it form} clouds \citep{BP+99a} than to drive turbulence into them \citep{Iffrig_Henneb15}, as also suggested by the recent observational realization that most local MCs appear to be located on expanding supernova shells \citep{Zucker+22}. On the other hand, if the turbulence is generated by the gravitational contraction itself \citep{VS+98, Kles_Henneb10, Robertson_Gold12, Xu_Laz20}, then the turbulent-to-kinetic energy ratio appears to be close to a virial-like value, even though the system is undergoing collapse, and very far from equilibrium \citep{GG_VS20}.

Unfortunately, a certain amount of confusion between the TS and GHC models is apparent in the recent literature, and the time is ripe for a precise comparison of the two models. In this paper we attempt to present such a comparison, first defining various fundamental terms precisely (Sec.\ \ref{sec:defs}) and then the main premises of each model (Sec.\ \ref{sec:premises}), to then discuss how various MC properties are interpreted in each scenario (Sec.\ \ref{sec:interpretation}). Next, we discuss their common ingredients and differences (Sec.\ \ref{sec:sim_diff}), to then suggest various possible observational and numerical tests for discerning which one, or combination of the two, is operating in molecular clouds (Sec.\ \ref{sec:tests}). Finally, we discuss other models that are related to TS and GHC, and share various of the features of each one (Sec.\ \ref{sec:disc}). Finally, we present a summary and our main conclusions in Sec.\ \ref{sec:concls}.
%In particular, a scenario that has received much attention in recent years is that of cloud-cloud collisions (CCCs) for the formation of high-mass star-forming regions \citep[e.g.,] [] {Fukui+18}. However, since this is not a model for the full structure and evolution of MCs, but rather just a scenario for the formation of massive regions, we only discuss it within the context of the TS and GHC models (Sec.\ \ref{sec:CCCs}).

\section{Definitions and precisions} \label{sec:defs}

\subsection{On the nature of turbulence} \label{sec:turb}

Turbulence is a chaotic flow regime containing a multitude of {\it coherent} structures (i.e., structures moving as a single unit, often referred to as {\it eddies}, or {\it vortices}) over a wide range of spatial scales, among which there is a {\it cascade} of energy from larger to smaller scales, such that the motions at smaller scales are fed by the larger scales. 

It should be noted, however, that the strict interpretation of {\it coherent structure} in turbulence, deriving from the incompressible case, does {\it not} refer to entire fluid patches moving ballistically through space as a single unit, but rather to regions of a given scale size across which there are well defined velocity {\it differences}. The archetypal incompressible coherent structure is the {\it eddy}, a vortex, across which there is a (transverse) velocity difference $v_\ell$, where $\ell$ represents the size (diameter) of the vortex. The equivalent coherent ``unit'' in the compressible case is known as a {\it shocklet} and corresponds to a {\it longitudinal} velocity difference across the scale $\ell$ \citep[see, e.g.,] [Sec. 13.5.1]{Lesieur08}. Therefore, the coherent turbulent structures should not be considered as objects moving ballistically (all at the {\it same} velocity), but rather as clearly identifiable regions where a particular main kind of motion (rotation, shear, compression or expansion) is happening.

Turbulence is a strongly dissipative process, and so energy must be continuously injected at the large-scale end of this hierarchy (referred to as the {\it injection} or {\it energy-containing scale}) of eddies in order to maintain a stationary state. The energy is dissipated at the small-scale end of the hierarchy, called the {\it dissipation scale}. The facts that the eddies are coherent structures and that the largest velocity differences occur at the largest scales imply that {\it the turbulent ram pressure cannot be isotropic}. The isotropic nature of thermal pressure arises from the availability of an enormous number of molecules traveling in all possible directions within a gas parcel. However, in the case of turbulence, the largest velocity contribution at the largest scale in a turbulent parcel is due to the single parcel-scale eddy that fits within the parcel, and therefore there is just one velocity vector associated to this eddy. A terrestrial analogy is that this is why strong atmospheric turbulence shakes aircraft rather than just increasing the pressure on the fuselage.

%This is because, for any scale considered, the dominant motion will be at that same scale, and so there is only {\it one} eddy at that scale that fits within the structure. Therefore, there is a well-defined velocity pattern, and averaging over a multitude of eddies (analogously to averaging over an ensemble of molecules) to 

Note that this is contrary to the extremely widespread assumption \citep[which dates back to] [] {Zuck_Evans74} that turbulence acts essentially as an additional temperature, consisting of small-scale motions that provide an isotropic pressure, which however depends on the scale considered. This is actually impossible to accomplish: if the motions are restricted to the small scales, their average over regions larger than that cannot depend on the size of the region considered. This is exemplified by the case of the thermal motions, for which, once a region is large enough for a statistical mechanical average to be valid, the value of the temperature does not depend on the size of the averaging region, unless the temperature itself varies with position. If turbulence were restricted to the small scales and were analogous to an additional temperature, then the property that velocity dispersion increases with the parcel size would be equivalent to temperature increasing with the size of the region over which it is measured. 

\begin{figure}
\centering
\includegraphics[width=1.\linewidth] {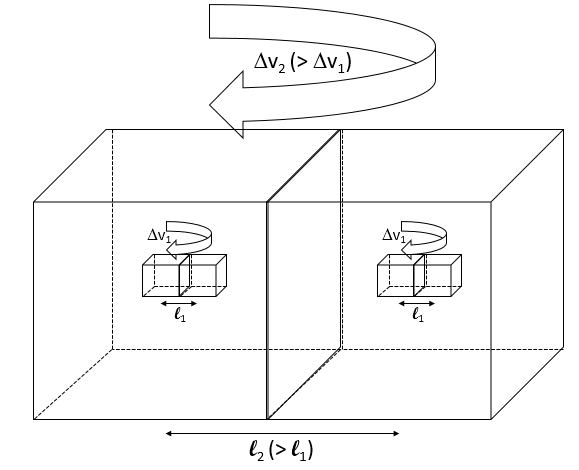}
\caption{A two-level schematic illustration of the multi-scale nature of the velocity differences in a turbulent flow. The ``eddies'' are represented as pairs of blocks rotating about each other. The size scale of the small-scale eddies is $\ell_1$, and the velocity difference across them is denoted $\Delta v_1$. The size of the large-scale eddy is $\ell_2$, and the velocity difference across it is $\Delta v_2$, and satisfies $\Delta v_2 > \Delta v_1$. In this sense, the small-scale eddies ``ride'' over the large-scale ones. For clarity, we have depicted the velocity differences as transverse (rotational), but the sketch applies equally to longitudinal (compressive) velocity differences.}
\label{fig:multi_scale_turb}
\end{figure}

Rather, this property requires that larger velocity differences occur across larger distances. This is illustrated schematically in Fig.\ \ref{fig:multi_scale_turb}, which shows a two-level hierarchical ``cloud'' of size $\ell_2$ consisting of two ``blocks'', among which there is a velocity difference $\Delta v_2$. In turn, the blocks are made up of smaller-scale similar systems, of size $\ell_1 < \ell_2$, and whose building blocks have a velocity difference $\Delta v_1 < \Delta v_2$. It is then clear that for such a system, when the velocity dispersion is measured over regions of size  $\sim \ell_1$, it will turn out to be of the order of $\Delta v_1$, while if measured over regions of size $\ell_2$, it will turn out $\sim \Delta v_2$. Only such a hierarchical system can allow for a scale-dependent velocity dispersion. But it precludes the large-scale ram pressure from being isotropic.
%two-level hierarchical structure with this property.  of a given size scale, the dominant velocity component is that at the scale of the object itself, and possesses a well-defined direction. Therefore, measurements of the turbulent velocity dispersion up to a certain scale will be dominated by velocity differences along that scale, allowing the dispersion to increase with region size.
%it cannot generate an {\it isotropic} pressure capable of providing support to the region similar to that provided by thermal pressure. 
Instead, the system-scale motions involve large-scale distortions (compression, shear, expansion or rotation) of the cloud or clump.

%In this case, when the dispersion is computed over small-scale regions, the large-scale components are effectively filtered out. This shows that the presence of large-scale motions 

%precludes, however, the possibility of turbulence being restricted to scales much smaller than the size of the region being considered, and of it providing an isotropic pressure.}

\subsection{On the intended meaning of ``Global'' in GHC. } \label{sec:global}

A term that has been the source of some confusion is that of ``global'' in the GHC model. By {\it global} it is meant that gravitational contraction is not restricted to occur only at the smallest scales ($\lesssim 0.1$ pc) in the clouds, but rather it can occur on scales up to the cloud itself \citep[e.g.,] [] {VS+07, VS+09, Heitsch+08a, Heitsch+08b}. In particular, anisotropic cloud-scale gravitational contraction was proposed by \citet[] [hereafter, GV14] {GV14} to be a feasible mechanism for the formation of long filamentary structures (see Fig. \ref{fig:sketch} and the left panel of Fig.\ \ref{fig:HFS_num}). 

However, this does not necessarily imply that the {\it entire} mass of the cloud must be undergoing gravitational contraction at all times, nor that the contraction is measurable in all directions. Concerning the former point, colliding-flow simulations of cloud formation show that the low-density outskirts of the clouds are often expelled by the very compression that forms them \citep[e.g.,] [] {VS+07, VS+11}. More recently, \citet{AV22} have shown that clumps (initially defined in three-dimensional space as connected regions above some threshold density) always lose some part of their mass when followed Lagrangianly over time, independently of the threshold used, as a consequence of angular momentum redistribution. Nevertheless, a substantial fraction of their mass still undergoes gravitational contraction. 

Concerning the direction of contraction, the collapse of an elongated structure leads to a clear collapse along the short axis in the standard free-fall timescale, while the longitudinal collapse takes a much longer timescale \citep[e.g., ][]{Toala+12, Pon+12}, roughly on a timescale comparable to the free-fall time corresponding to the mean density of the sphere that contains the same mass as the elongated structure and circumscribes it. Such a larger timescale makes it much harder to recognize the nature of a global contraction and, more importantly, has relevant consequences in reducing the efficiency of star formation (Zavala-Molina et al., in prep).

Finally, the expansion of an HII region can produce simultaneously dense, locally gravitationally unstable regions that will proceed to collapse, regardless of them being pushed away from each other by the HII region---the well-known ``collect and collapse'' effect \citep{Elm_Lada77}.

\subsubsection{Can molecular clouds be stabilized by turbulence?} \label{sec:can_turb_stabilize_MCs?}

The fundamental consideration in the context of GHC is that, containing up to thousands of Jeans masses, molecular clouds are strongly Jeans-unstable, and are thus subject to a strong, {\it continuous}, self-gravitational pull. Turbulence may, under some circumstances, oppose the gravitational pull, but this needs to be considered carefully. As discussed by \citet{Ballesteros-Paredes06}, not all turbulence is available for support. At any scale, turbulence may contribute simultaneously to smaller-scale structure formation and larger-scale structure support. Moreover, not all nonthermal motions correspond to turbulence. Instead, a significant fraction of the kinetic energy may correspond to infall motions that do not provide support. We can identify three main kinds of turbulent energy injection that can be applied to MCs: 1) from the cloud assembly process itself, 2) from external sources like Type Ia supernovae (SNe), and 3) from internal sources, like outflows, {\sc Hii} regions and possibly Type II SNe \citep[although even these usually explode in regions previously evacuated by \hii\ regions; e.g.,] [] {Haid+19}. Driving from the cloud assembly process itself \citep[e.g.,] [] {Koy_Inu02, Kles_Henneb10} is, in general, insufficient to prevent the gravitational contraction of the cloud once it reaches its Jeans mass \citep{VS+07, Heitsch+08a, IbanezM+16, GG_VS20}. Driving from {\it external} sources can either {\it form} new clouds \citep[e.g.,] [] {BP+99a, IbanezM+17, Zucker+22} through compressions similar to those modeled in colliding-flow simulations, or else shear and shred pre-existing clouds, as in the simulations by \citet{Padoan+16}. Finally, driving from {\it internal} sources rapidly destroys or disperses most of the clouds' mass \citep[e.g.] [] {Colin+13, Peters+17, Grudic+22}. In either case (internal or external driving), the stellar feedback is {\it impulsive} (momentary and intermittent in time), while self-gravity operates {\it continuously} in time.
%{\color{magenta} no deberíamos mencionar alfvén waves from newborn massive stars\citet[][(Offner y Liu 18)]{Offner_Liu18} EVS: no lo puse, pues me parece medio fumada la idea, y adem[as me obligaria a refrasear mucho el texto aqui.}

In any case, neither of the above energy injection mechanisms provides a means of {\it stabilizing} the clouds against the systematic and continuous action of self-gravity, and so the clouds are likely to remain {\it out of equilibrium}. Collapse-driven turbulence simply appears to be dissipated too rapidly to halt or significantly delay the collapse \citep{GG_VS20}. External driving may perturb the clouds periodically while they are contracting, sometimes ``chipping of'' the outer parts of the clouds, but also generating compressions that form new clouds or enhance already-ongoing collapses. Finally, internal driving eventually destroys the clouds, at least locally, except possibly in the case that the clouds are extremely massive \citep[e.g.,] [and references therein] {Dale15}. In very massive complexes, the internal driving can perhaps only interrupt the current SF episode, push the gas sideways, and initiate another episode at a nearby location. 

\subsubsection{What scales can undergo gravitational contraction in molecular clouds?}

In conclusion, the intended meaning of the ``global'' qualifier under GHC is that {\it the infall motions do not need to be constrained to the smallest scales (cores, $\ell \lesssim 0.1$ pc), but can extend to the size of the entire molecular complexes}. In particular, GHC proposes that even apparently unbound structures, such as very long filaments, may be the resul of cloud-scale transverse gravitational contraction (GV14). On the other hand, it specifically does {\it not} imply that the totality of the molecular material is undergoing infall at all times. Due to the impulsive nature of the feedback, external driving will sometimes ``strip off'' gas from the outer parts of the clouds, and low-energy feedback may set up diverging flows amid the global infall. Finally, strong internal feedback must eventually disrupt the clouds, shredding them to pieces that may be flying apart, and thus constituting truly unbound molecular gas \citep[e.g.,] [] {Evans+21}.

\subsection{GHC in the Galactic context} \label{sec:GHC_galactic}

\begin{figure*}
\centering
\includegraphics[width=1.\linewidth] {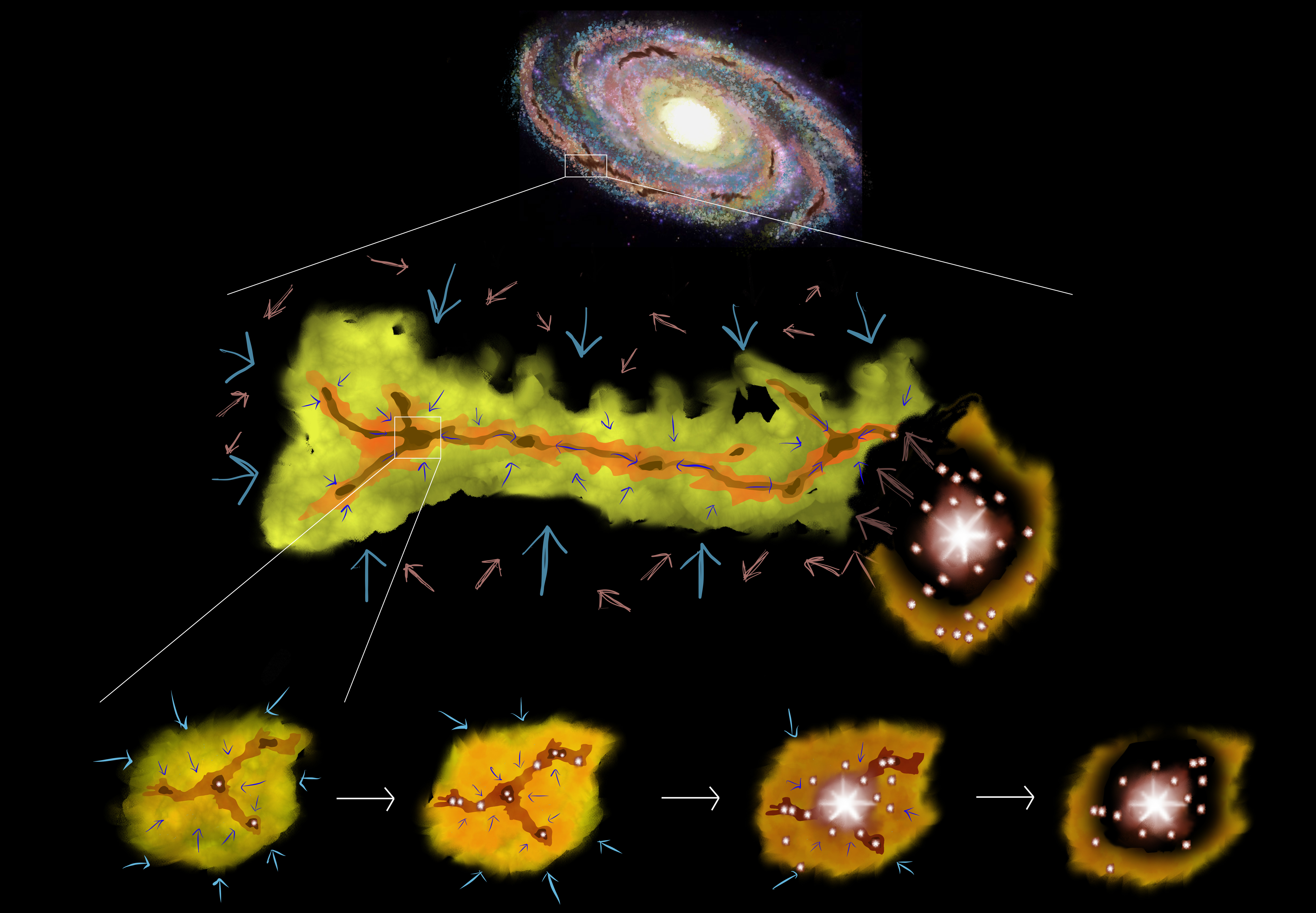}
\caption{Schematic illustration of the GHC scenario from spiral-arm scales down to molecular cloud scales, across time. Colored arrows denote the motion of gas, with pink indicating inertial motions and blue indicating gravitational motions. The sketch combines spatial zoom-ins from top to bottom and temporal evolution from left to right in the bottom row (indicated by the white arrows). See Sec.\ \ref{sec:GHC_galactic} for a description of the various stages.}
\label{fig:sketch}
\end{figure*}

In view of the previous discussion, it is also appropriate to describe GHC in the Galactic context, as previously it has only been discussed up to the scale of giant molecular complexes. This is schematically depicted in Fig.\ \ref{fig:sketch}, which from top to bottom shows the scale hierarchy of gravitationally formed structures, and from left to right in the bottom row illustrates the temporal evolution of a star-forming hub-filament system (HFS).

At the largest scales ($\sim$ several kpc), gas is gravitationally compressed as it passes through spiral arms, although in this case the relevant gravitational instability criterion is most likely the Toomre criterion for the combined gravity of stars and gas \citep[e.g.,] [] {Rafikov01, Li+05}. This mechanism produces large, spiral, cold gas complexes, aided by thermal instability. As the gas cools down and becomes denser, it begins to simultaneously become molecular, Jeans unstable, and magnetically supercritical \citep{HBB01}. The condensation from the WNM to the CNM causes a precipitous drop in the Jeans mass, allowing the newly formed MCs to rapidly acquire a large number of Jeans masses.

Due to various instabilities, the condensed gas develops moderately-supersonic turbulence \citep[e.g.,] [] {Walder_Folini00, Koy_Inu02, Audit_Henneb05, Heitsch+05, Heitsch+06, VS+06}, which produces nonlinear density fluctuations in the CNM. The presence of these fluctuations causes the large-scale gravitational contraction to be extremely non-homologous \citep[i.e., non-monolithic, with the densest parts culminating their collapse earlier;] [] {Larson69}, and therefore a continuous {\it gravity-driven accretion flow} develops from low- to high-density regions \citep[e.g.,] [] {Burk_Hart13}. Moreover, this flow is highly chaotic and anisotropic, since the large Jeans mass content implies that the gas behaves in a nearly pressureless manner (GV14), causing the gravitational contraction to amplify anisotropies \citep{Lin+65}, and so {\it the ``centers of collapse'' may actually be surfaces or lines rather single points}. This process generates dense filamentary structures, that accrete from the bulk of the molecular cloud, and increase their central density (GV14). However, local density fluctuations within the filaments grow faster than the rest, becoming dense cores first, and eventually massive hubs. The latter are massive and dense enough to drive a longitudinal flow that prevents the formation of strong shocks at the filament axis, since the radial inflow is smoothly changed into a longitudinal one by the gravitational pull of the faster-collapsing hubs in the filaments \citep{GV14, Gomez+18, Naranjo+22}.

Finally, the bottom row of Fig.\ \ref{fig:sketch} illustrates, from left to right, the process of growth of both the stellar and gaseous mass in a hub. This process continues until sufficiently massive stars form at the hub \citep{VS+24} as to disrupt the local star-forming flow, ending the local star formation episode at an early stage, thus keeping the global SFR low, while simultaneously compressing the preexisting neighboring molecular gas, to form new MCs and new star formation episodes,. These compressions can be considered the ``aftershocks'' of the initial ``earthquake'' (the infall of gas into the spiral potential well), and might not end until the gas leaves the spiral arm. This may be the case, under the GHC scenario, of most MCs in the Solar Neighborhood, which appear to lie at the intersections of supershells \citep[e.g.,] [] {Zucker+22}, and to be undergoing their own collapse process \citep{Zhou+24b}.

%effect of a massive star pushing aside some of the gas in the large complex and producing new star-forming clouds in the compressed shell, {\cyan as suggested by the observational work of \citet{Zhou+23}}. 
In summary, in the GHC scenario, the clouds are in general out of equilibrium \citep{IbanezM+17}, in such a way that, when they are not being dispersed, they are attempting to undergo global collapse amidst intermittent perturbations from the outside.
%{\it mostly} dominated by self-gravity until they are destroyed by internal feedback.

\begin{figure*}
\centering
\includegraphics[width=0.9\linewidth] {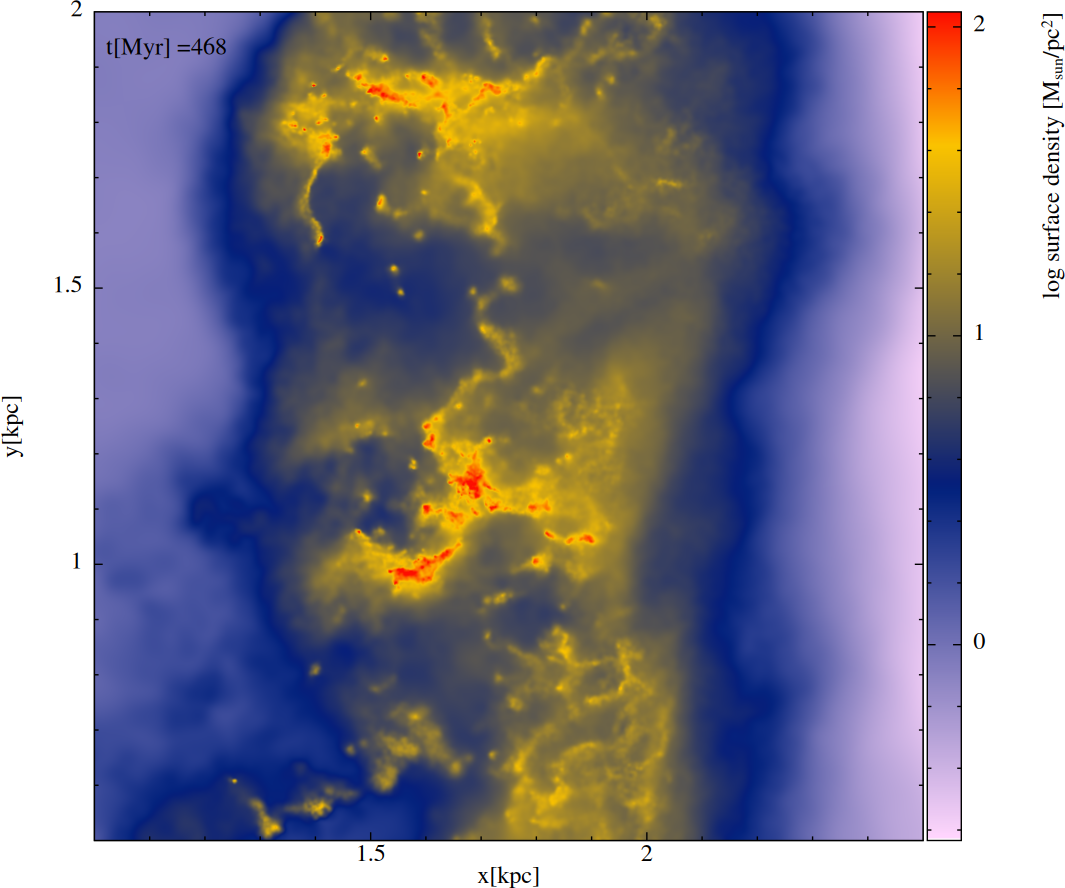}
\caption{Column density field in a 1.5-by-1.5 kpc subregion of the numerical simulation described in Sec.\ \ref{sec:GHC_galactic} and Appendix \ref{app:simulation}, showing various clouds and cavities carved by SN explosions. In the electronic version, this figure is an animation showing the evolution of the clouds, and the fact that they are never in equilibrium, but rather, when they are not undergoing dispersal, they are attempting to collapse, amidst a continuous bombarding by the SN explosions, which some times {\it form} the clouds and others disperse them.}
\label{fig:straight-arm_sim}
\end{figure*}

To illustrate this process, in Fig.\ \ref{fig:straight-arm_sim} (available as an animation in the electronic edition) we show an excerpt from a numerical simulation following the flow of initially diffuse gas ($n = 0.05 \pcc$)\footnote{Note that this density is actually almost an order of magnitude lower than the density typically associated with the warm atomic gas in the Solar Neighbourhood, but this roughly compensates for the fact that the simulation does not include galactic rotation, and therefore the gas is more gravitationally unstable than if it did.} as it enters a gravitational potential well representing a spiral arm, except that the potential in this case is straight rather than spiral for simplicity. The simulation is described in detail in Appendix \ref{app:simulation}, but here we just mention that it represents a region of 4 kpc per side, into which warm diffuse gas enters through the $x=0$ boundary. The gas is subject to its self-gravity as well as to an ``arm'' gravitational potential that extends over the full box length along the $y$ direction and is placed at the middle of the $x$ and $z$ dimensions. The gas is compressed as it falls into the external potential, forming cold dense clouds, which eventually begin to undergo gravitational collapse and to form sink particles that produce supernova (SN) explosions. The latter in turn rapidly disperse the remaining parts of the clouds that have not produced sink particles yet. These regroup somewhere else and rapidly begin to undergo gravitational collapse again, repeating the process until the gas leaves the arm potential well.

This simulation clearly illustrates that dense structures of sizes up to $\sim 100$ pc are never ``supported'' nor maintained in near equilibrium, but instead are either in the process of being dispersed, or else attempting to collapse. This is the essential premise of the GHC scenario. 
%Part of the cloud is infalling while part of the cloud is being dispersed, but no part is in equilibrium. 
This mechanism of alternation between collapse and dispersal is also apparent in simulations with feedback and self-gravity by other groups \citep[e.g.,] [] {Kim+21, Kim+23, Grudic+22}, although it is not always recognized that this is consistent with the GHC model, and instead it is often described as TS. An equilibrium description may be valid when large enough temporal or spatial scales are considered, so that averages over entire cloud ensembles or over many cycles of collapse/dispersal are performed, but not on the spatial and temporal scales of individual cloud evolution. An estimate of the minimum size scales over which the averaging must be performed may be the region size required for measurements of the depletion time $\tau_{\rm dep} \equiv M_{\rm gas}/\dot M_*$ (measured as the ratio of the H$\alpha$-to-CO emission) performed around CO (gas) emission peaks to coincide with measurements performed around H$\alpha$ emission (SFR) peaks. This scale has been determined to be $\gtrsim 300$ pc \citep{Schruba+10}.

A similar mechanism is also observed at smaller scales in the interiors of MCs, and corresponds to the ``clump fed'' accretion onto YSOs observed in the simulations of \citet{Wang+10},  who concluded that the feedback from bipolar outflows was able to partially oppose, but not stop, accretion from the ``clump scale'' ($\sim 1$ pc) to the ``core scale'' ($\sim 0.1$ pc). Instead, our simulation exhibits this mechanism at scales of up to $\sim 100$ pc, suggesting that it operates on all scales within MCs. This is also illustrated schematically in Fig.\ \ref{fig:sketch}, which shows accretion at all scales from the molecular complex scale down to the core and hub scale, and, ultimately, to the protostellar accretion disk scale.

\subsubsection{Largest scales of GHC applicability} \label{sec:upp_end_GHC}

One final remark, however, is that GHC, as proposed in \citet{VS+19}, relies exclusively on the thermal support, described by the Jeans criterion. However, on large enough scales, the Galactic differential rotation also provides support, stabilizing the {\it largest} scales. This support definitely needs to be taken into account \citep[e.g.,] [] {Meidt+18}. A simple estimate of the maximum spatial scales of applicability of GHC can be made by considering the Toomre analysis,
%\Green{. For a given $Q<1$ parameter, the range of unstable wavenumbers $k$ is given by
%%
%\begin{equation}
%    k = 2 k_\text{crit} \left[ \frac{1 \pm (1-Q^2)^{1/2}}{Q^2}\right],
%\end{equation}
%%
%where $k_\text{crit}=\kappa^2/( \pi G \Sigma)$ is the critical wavenumber for a pressureless, zero-thickness disk \citep{B&T08}, $\kappa \approx 39.1 \kms {\rm kpc}^{-1}$ is the Galactic epicyclic frequency \citep{Li+19}, and $\Sigma$ is the gas surface density. The small $k$ limit corresponds to the largest gravitationally unstable scale, and thus it provides an estimation of the scale at which GHC is no longer applicable. For
%}
according to which, the first wavelength to go unstable when $Q \lesssim 1$ is
\begin{equation}
    \lambda_{\rm d} = \frac{2 \pi^2 G \Sigma}{\kappa^2} = 55.6~{\rm pc} 
    \left[\frac{\Sigma}{\Msun {\rm pc}^{-2}} \right],
    \label{eq:lambda_unst_Toomre}
\end{equation}
where $\Sigma$ is the gas surface density, and $\kappa \approx 39.1 \kms {\rm kpc}^{-1}$ is the Galactic epicyclic frequency \citep{Li+19}. This expression shows that 
molecular gas, which typically has surface densities $\Sigma > 10 \Msun$ pc$^{-2}$, is not stabilized by differential rotation in the Galaxy for scales smaller than $\sim 550$ pc. Therefore, we expect GHC to be applicable up to scales of a few hundred parsecs.

\subsection{Collapse} \label{sec:collapse}

Another term that has turned out to be prone to confusion is that of ``collapse''. In \citet{VS+19} and throughout the present paper, we use it to denote the {\it process} (extended over time) of gravitational contraction, and the associated {\it flow} that generates accretion onto each density level of the hierarchy, in particular onto and along filamentary structures within MCs (see Sec.\ \ref{sec:hier_coll}). This is to be distinguished from the instantaneous {\it event} of the formation of one or many collapsed objects (in practice, protostellar objects, and, mathematically, singularities in the flow). 
In this same line of argument, the collapse flow described in the GHC model {\it does not} correspond to a monolithic collapse of the cloud, as has been interpreted by some authors \citep[e.g.,] [] {Krumholz_McKee20}. Instead, it involves a continuous {\it conveyor belt} \citep{Longmore+14} accretion flow funneling material from low to high densities (see discussion in Sec. \ref{sec:monolithic_GC}).  

\subsection{Hierarchical, asynchronous collapse} \label{sec:hier_coll}

The term {\it hierarchical} in the GHC context refers to the fact that gravitational contraction is in general non-homologous and induces fragmentation, as proposed by \citet{Hoyle53}. The non-homologous nature causes the denser parts of the collapsing cloud to reach the singularity {\it sooner} than the less dense parts. In addition, the (moderately supersonic) turbulence produces multiple density enhancements which can begin to undergo their own gravitational contraction, if their masses exceed the (temporally decreasing) mean Jeans mass in the medium. This generates multiple instances of collapse at multiple locations and at different times and rates, depending on the locations and local masses and densities of the fluctuations, with lower-mass regions starting their collapses later because they can only do so when the mean Jeans mass in the cloud has decreased to the value of their own masses \citep{VS+19}. We refer to this as an {\it asynchronous} collapse. Moreover, the contraction amplifies anisotropies, and thus becomes extremely anisotropic, forming filamentary structures (see Sec.\ \ref{sec:GHC_def}.vi below). Therefore, far from being a single, monolithic, isolated event, the collapse forms a hierarchy of objects, such as clumps, filaments, cores, and, finally, stars. The collapse flow also drives accretion onto each of these levels of the hierarchy.

\section{Premises of the models} \label{sec:premises}

\subsection{The turbulent support model} \label{sec:TS_def}

The fundamental premises of the TS model are:

\begin{enumerate}

\item MCs are {\it objects} \citep[i.e., structures consisting of roughly the same material throughout their evolution; see Sec.\ 3.1 of] [for the distinction between object-like and wave-like structures.] {VS+96} in approximate equilibrium, with roughly fixed masses, supported against global collapse by strongly supersonic turbulence \citep[e.g.,] [] {Zuck_Evans74, Tan+06} and/or a strong magnetic field \citep[e.g.,] [and references therein] {Shu+87, Mouschovias91}. The near-equilibrium implies long lifetimes and quasi-stationary (non-evolving) conditions.

\item The nonthermal component of the linewidth is interpreted as a manifestation of the supersonic turbulence providing support against self-gravity.

\item The near-virial velocity dispersions observed in MCs and their substructures \citep[e.g.,] [] {Larson81, Heyer+09} indicate that turbulent kinetic energy approximately balances the structure's gravitational energy, allowing near-virial equilibrium.

\item The collapse criterion is given by the ``turbulent Jeans mass'', $\MJt$, based on the total (thermal+nonthermal) velocity dispersion, rather than on just the thermal one. This mass must be exceeded in order for collapse to occur.

\item All the substructures of MCs (filaments, clumps, and cores) are the result of supersonic turbulent compressions \citep[e.g.,] [] {Padoan+01}.

\item Only the smallest scales (MC cores, with densities $n > 10^4\, \pcc$ and sizes $\Lcal \lesssim 0.1$ pc) are able to collapse, if they manage to locally exceed their own Jeans mass upon the external compression \citep[e.g.,] [] {Pad_Nord02, VS+03, Krum_McKee05}.

\item In theoretical models of the control of the SFR under the TS scenario, the SFR is a stationary property of the clouds that depends on the main turbulent parameters, namely the virial parameter $\av$ (defined in eq.\ \eqref{eq:Larson_vir} below), the sonic Mach number of the turbulence, the turbulent forcing parameter $b$ (a measure of the energy fraction in compressible turbulent modes, as opposed to rotational modes), and the plasma $\beta$ or, equivalently, the Alfv\'enic Mach number \citep[e.g.,] [] {Federrath_Klessen12}. Assuming that the turbulence virializes every scale, \citet{Krumholz+12} have suggested that ultimately the SFR depends only on the gas number density, in such a way as to maintain a constant value of the {\it star formation efficiency over a free-fall time}, the fraction of a cloud's mass that is turned to stars over the free-fall time for the mean cloud's density, of order $\epsff \sim 1\%$. The low fraction of gas mass converted to stars is attributed to turbulent support \citep{Krum_McKee05}, which prevents global collapse at the cloud scale ($\gtrsim 1$ pc) while inducing local compressions that may become locally unstable and collapse, at the core scale ($\lesssim 0.1$ pc) \citep{VS+00, VS+03, MacL_Kles04}.

\end{enumerate}

\subsection{The global hierarchical collapse model} \label{sec:GHC_def}

On the other hand, the GHC model's fundamental premises are:

\begin{enumerate}

\item MCs and their substructures are approximately stationary {\it flow features}, i.e., they are the {\it loci} through which the gas flows towards the (multiple) troughs of the gravitational potential, analogous to the {\it flame} in a burning candle: the gas continues to flow through it, but the location of the flame remains stationary. They are not {\it systematically} supported against collapse at any particular scale by turbulence nor magnetic fields, and the gas flowing through them is in general {\it out of dynamical equilibrium}. 

\item The nothermal component of the linewidth is interpreted as a mixture of infall and turbulent motions, in which the infall component becomes progressively dominant as the column density of the objects increases \citep[eq.\ (8) and Fig.\ 2 of] [] {VS+19}.

\item The near-virial velocity dispersions observed in MCs and their substructures \citep[e.g.,] [] {Larson81, Heyer+09} are the manifestation that they are in the process of gravitational contraction, with the kinetic energy consisting predominantly of infall motions. Low-column density or low-mass clouds that appear strongly supervirial are either in an assembly stage, being dominated by external compressions, or in late stages of dispersal, dominated by energy injection from stellar feedback.

\item The collapse criterion is given by the standard thermal (rather than the turbulent) Jeans mass, since the kinetic energy in gravity-dominated objects corresponds mainly to infall motions that do not oppose the collapse, but rather are a consequence of it.%\footnote{{\mgt In any case, note that, as pointed out in Sec.\ \ref{sec:turb}, the truly turbulent motions are predominantly at large scales, and therefore they cannot provide hydrostatic support for a cloud. Instead, they cause a large-scale {\it distortion} of the cloud that may prevent it from collapsing (in the case of shear or dispersal, for example), but this does not render the cloud gravitationally stable. As shown in the simulation presented in Sec.\ \ref{sec:global}, a cloud can be globally Jeans-unstable and thus be attempting to collapse, searching to ``percolate'' through the turbulent motions that work to distort it.}}

\item All the substructures of MCs (filaments, clumps, and cores) tend to grow due to multi-scale (i.e., ``global''), anisotropic gravitational contraction before massive stars begin to form \citep{Burk_Hart04, Heitsch+09, BP+11, GV14}.

\item \label{sec:item_non-monolithic} The gravitational contraction, however, is far from monolithic and radial, but is rather a strongly chaotic (nonlinear), anisotropic \citep[from volumes to surfaces to filaments to cores;] [] {Lin+65}, and hierarchical process, consisting of small-scale collapses within larger-scale ones, often in a conveyor-belt fashion. The smallest, densest regions start their own collapse later than the large scales, because they need to wait until the Jeans mass has shrunk to their own value. However, they terminate their collapse earlier, because of their significantly shorter free-fall time \citep{VS+19}. 

\item In theoretical models of the control of the SFR under the GHC scenario \citep[e.g.,] [] {ZA+12, ZA_VS14, VS+18, VS+19, Camacho+20}, the SFR is an intrinsically {\it evolving} property of the clouds, essentially governed by the evolution of the collapse flow, and later by stellar feedback. During the early evolutionary stages of MCs, the SFR first increases over time as the clouds evolve to higher masses and dense gas mass fractions by accretion from their parent structures, due to the large-scale gravitational contraction flow. At later stages, the SFR becomes large enough to form massive stars that disrupt the clouds. In the simplest model of GHC, including only (constant) accretion, large- and small-scale collapse, and photoionising feedback, and assuming clouds start at densities typical of the CNM, a cloud's evolution is completely specified by two independent parameters, for example its age and its mass at that age \citep{ZA+12, ZA_VS14, VS+18}, or its age and the accretion rate \citep{Camacho+20}. Two parameters are necessary because clouds increase their mass as they age, and so their mass at a given time depends both on their age and on the accretion rate onto them. The density-SFR correlation is a byproduct of the gravitational contraction \citep{VS+18, Camacho+20}.

\item The growth rate of the mass, density, and SFR begins to decrease as the stellar feedback increases, which eventually reverses the growth, causing the cloud to begin losing mass and decreasing its SFR, until the local episode of star formation is terminated \citep{ZA+12, Colin+13, ZA_VS14, VS+18}. New episodes can be triggered in the neighbouring compressed gas, though.

\end{enumerate}

\section{Interpretation of molecular cloud properties} \label{sec:interpretation}

Under the above sets of fundamental premises, each model interprets various properties of MCs differently. We here list a number of these properties and their respective interpretation within the context of each model. For economy, in what follows, we will use the term ``clump'' in a generic way, denoting anything from a GMC to a dense core, %\gil{Would you agree to describe a clump as ``any region significantly denser than its surroundings''?} 
thus generically describing any region that is significantly denser than its surroundings, except when the distinction becomes necessary. 

\subsection{Larson's relations and their generalization} \label{sec:Larson_rels}

Molecular clouds have been known for several decades \citep{Larson81} to obey two scaling relations, one between their mean density and radius, 
\begin{equation}
\langle n \rangle \approx 1.7 \times 10^3 \pcc \left(\frac{R} {{\rm pc}} \right)^{-1},
\label{eq:Larson_n_L}
\end{equation}
which implies that they all have approximately the same column density, and the other between the clouds' velocity dispersion (or, equivalently, their linewidth) and their size. The currently accepted form of this relation is \citep[e.g.,] [] {Heyer_Brunt04}
\begin{equation}
\sigma_v \approx 1.5 \kms \left(\frac{R} {{\rm pc}} \right)^{1/2}.
\label{eq:Larson_sigma_L}
\end{equation}
Furthermore, as is also well known, taken together, these relations imply near equipartition between the nonthermal kinetic energy and the gravitational energy, so that the {\it virial parameter} \citep{Bertoldi_McKee92}, defined as
\begin{equation}
\av \equiv \frac{2\Ek}{|\Eg|} \approx \frac{5 \sigma_v^2 R}{GM} = \frac{5 \sigma_v^2} {\pi G \Sigma R},
\label{eq:Larson_vir}
\end{equation}
is constant and of order unity. Here, $M$ is the cloud's mass, $\Sigma \approx M/\pi R^2$ is the column density, and $R$ is some measure of the cloud's size. An equivalent statement is that the so-called {\it Larson ratio}, $\Lcal \equiv \sigma_v/R^{1/2}$, satisfies \citep{Keto_Myers86, Heyer+09}
 \begin{equation}
 \Lcal = \left(\frac{\pi \av G \Sigma} {5}\right)^{1/2}.
 \label{eq:Larson_ratio}
 \end{equation}
In what follows, we refer to this scaling as the {\it Keto-Heyer} (or KH) relation.

\begin{enumerate}

\item {\it Interpretation in TS:} 

The linewidth-size relation is the manifestation of the expected scaling between the velocity dispersion and the eddy size for a turbulent medium, with a specific (i.e., per unit mass) kinetic energy spectrum $E(k) \propto k^{-q}$, where $k$ is the wavenumber, and $q$ is the ``spectral slope''. Such a spectrum implies a velocity dispersion-size scaling relation of the form $\sigma_v \propto \ell^\eta$, where $\ell \sim 1/k$ is the size scale and $\eta= (q-1)/2 = 1/3$ for incompressible Kolmogorov turbulence ($q = 5/3$), while $\eta = 1/2$ for {\it Burgers'} pressureless turbulence ($q = 2$). Strongly supersonic turbulent simulations typically take values of the spectral slope $q \lesssim 2$ \citep[e.g.,] [] {Kritsuk+07}, implying $\eta \lesssim 1/2$, in agreement with observational determinations \citep[e.g.,] [] {Larson81, Solomon+87, Heyer_Brunt04}. Therefore, the observed velocity dispersion-size scaling, eq.\ \eqref{eq:Larson_sigma_L}, is interpreted as the manifestation of the turbulent energy spectrum of strongly supersonic turbulence. 

In addition, the density-size relation, eq.\ \eqref{eq:Larson_n_L} is interpreted as a characteristic feature of MCs \citep[e.g.,] [] {Lombardi+10}, which all have similar column densities, so that their volume density scales inversely with their size. Since this is the column density for which the turbulent velocity dispersion equals the virial velocity \citep{Larson81}, MCs are interpreted in TS as structures in which supersonic turbulence is in near-virial balance with their self-gravity.

\bigskip
\item {\it Interpretation in GHC:} 

The %main physical property of the clouds is the 
near equipartition between the kinetic and gravitational energies of the clouds is interpreted in GHC not as virial equilibrium, for which $2 \Ek = |\Eg|$, but rather as near free-fall, for which $\Ek \approx |\Eg|$ \citep{BP+11}, and therefore, $\av \approx 2$. Here it is important to note that observational studies indeed seem to be more consistent with $\av \sim 2$ that with $\av \sim 1$ \citep[e.g.,] [] {Heyer+09, Field+11, Sun+18}.

The Larson density-size relation for molecular clouds, eq.\ \eqref{eq:Larson_n_L}, is interpreted only as an observational selection effect present in early observations \citep[e.g.,] [] {Solomon+87} due to the definition of MCs in terms of CO observations, which required at least a certain minimum column density in order to  allow for CO molecule formation. As shown by \citet{BP+12}, for power-law or lognormal column density PDFs, the mean column density of the clouds is very close to the threshold value, and therefore clouds defined by their detectability by a given tracer appear as all having similar column densities. More modern studies \citep[e.g.,] [] {Heyer+09, MivilleD+17} show that the clouds' column densities can span at least a few orders of magnitude, supporting the view that Larson's density-size relation arose simply from an observational selection effect. 

Nevertheless, MCs in general are found to satisfy eq.\ \eqref{eq:Larson_ratio}, which indicates near energy equipartition. Furthermore, structures with much higher column densities, such as massive star-forming clumps, continue along the same scaling as that of MCs \citep{BP+11, BP+18}. Therefore, relation \eqref{eq:Larson_ratio} should be considered as the generalization of the Larson velocity dispersion-size scaling when the density-size relation is relaxed, and applicable to all molecular structures, not just large, relatively-low column density molecular clouds. 

%Furthermore, near equipartition {\mgt between the kinetic and gravitational energies} is also consistent with free-fall of the clouds, since the difference between the free-fall and the virial velocities is only a factor of $\sqrt{2}$, and thus usually buried within the uncertainties \citep{BP+11}, making free-fall nearly indistinguishable from virial equilibrium through energy-budget indicators like $\av$ and $\Lcal$. Therefore, under GHC, the physical meaning of the observed near energy equipartition is that the motions are dominated by gravitational contraction. Nevertheless, large data collections tend to favor the free-fall value $\av =2$ from a statistical standpoint \citep[e.g.,][] {Sun+18}.

\subsection{Deviations from virial balance. The $\av$-$M$ and $\Lcal$-$\Sigma$ relations} \label{sec:devs_vir}

Although to zeroth order MCs and their substructures are in near equipartition between their gravitational and nonthermal kinetic energies \citep[$\Ek \sim \Eg$;] [] {Larson81, Heyer+09}, systematic deviations are observed, such that the clumps with the lowest-column densities and/or masses tend to appear significantly supervirial \citep[e.g.,] [] {Keto_Myers86, Kauffmann+13, Leroy+15, MivilleD+17, Traficante+18b}, while those with the largest column densities and/or masses often appear subvirial \citep[e.g.,] [] {Pillai+11, Kauffmann+13, Zhang+15_superJeans, MivilleD+17, Traficante+18b, Traficante+23}.

\begin{enumerate}

\item {\it Interpretation in the TS model}

\begin{enumerate}

\item As mentioned in Sec.\ \ref{sec:premises}, in the TS model, as well as in many observational works, the linewidth is interpreted exclusively as turbulence, since the structures are assumed to be in approximate virial equilibrium. Therefore, the velocity dispersion $\sigma_v$, and the associated virial parameter $\av$ and Larson ratio $\Lcal$ are cloud {\it parameters} (fixed for a given cloud), representing only the turbulent motions, which provide support against self gravity.

\item The supervirial nature of the low-column density (respectively, low-mass) clumps in the KH (resp.\ \aM) diagram implies that either they are over-supported by turbulence and in need of external pressure confinement to remain in equilibrium \citep[e.g.,] [] {Keto_Myers86, Bertoldi_McKee92, Field+11, Leroy+15}, or else they are gravitationally unbound \citep{Evans+21} and probably in the process of dispersal.

\item The sub-virial velocity dispersions often observed in high-column density clumps in the KH diagram or high-mass clumps in the \aM\ diagram indicate that the objects lack turbulent support and thus are ``on the verge'' of collapse, or else they are supported by other means, such as the magnetic field \citep[e.g.,] [] {Tan+13, SanchezM+13, Zhang+15_superJeans, Kong+18, Scibelli+23}.

\item \citet[] [Sec.\ 3.2.2] {Krumholz+19} have proposed that $\av$ should decrease with cloud {\it radius} if the clouds obey Larson's linewidth-size relation. In this case, $\sigma_v \propto R^{1/2}$, and so their kinetic energy per unit mass, $e_{\rm k} \approx \sigma_v^2 \propto R$. On the other hand, their gravitational energy per unit mass is $e_{\rm g} \approx GM/R \propto R^2$, if ``one chooses a region close to the mean density'', so that $M \approx \langle \rho \rangle R^3$. Therefore, $\av \equiv 2 e_{\rm k}/e_{\rm g} \propto R^{-1}$. This prediction is consistent with the scaling $\av \propto R^{-1.13 \pm 0.10}$ found by \citet[] [see their Fig.\ 10] {Traficante+18b}.\footnote{Note, however, that assuming the same mean density for objects of smaller sizes ignores the fundamental fact that the relevant small-scale substructure of molecular clouds is denser than the mean. The interpretation by \citet{Krumholz+19} thus selects the low density ``chaff'' at small scales, rather than structures like filaments, clumps and cores, and thus is not applicable to them \citep{BP+24}.
}

In addition, this same set of assumptions implies that $M \propto R^3$, and therefore $\av \propto M^{-1/3}$, which is {\it qualitatively} consistent with the observed trend of decreasing $\av$ with $M$, although it should be noted that a wide range of scaling exponents $\beta$ is reported in observational works ($-0.99 \le \beta \le 0.12$) \citep[e.g.,] [] {Kauffmann+13, MivilleD+17, Traficante+18b, Traficante+23}, since uncertainties strongly affect the derived value of $\beta$ \citep{Kauffmann+13}, and therefore this exponent is not well observationally constrained.

%{\color{magenta} JBP (pochismos cortesía del corrector de hortographye): Respecto a los últimos dos párrafos, yo lo refrasearía different: el ejercicio de Krumholz está bien, pero es válido solo si la densidad es CST. Pero results que la densidad no es constant. Los cores densos tienen densidad mayor que la nube de CO. Piriod. Entonces, si te clavas en que la densiad es constant, si: las structures más grandes están ligadas, y las más chicas no. Pero esas chicas son chaff. El problems de clavarse con la pendiente es: (a) Ya Jens demostró que las incertidumbres juegan un papel importante en determinar la pendiente final de esta relación \citet{Kauffmann+13}). Yo no me atrevería a decir ``the typical scaling''. Por ejemplo, si te crees las 2 relaciones de Larson exactas, la pendiente debería ser cero (Larson81, figura 4) (b) Hoy día ya hay gente que ha comenzado a hacer thresholds de densidad volumétrica. Esos cores bien pueden dar una alfavir vs mass con pendiente de -1/3. ¿so what? Que haya cores que den otra pendiente en la relation alfavir-masa no prueba lo chueco del argument de Krumholz.}

\end{enumerate}

\item {\it Interpretation in the GHC model}

%{\cyan Traficante+23 mention that the alphavir vs Mass anticorrelation may arise because most massive clumps are more gravitationally bound and the gravitational potential overwhelms the local turbulence (Ballesteros-Paredes et al. 2011). THIS DOES NOT SEEM FULLY CONSISTENT WITH OUR EXPLANATION IN 3) SHOULD WE MENTION THIS?}

\begin{enumerate}

\item As mentioned in Sec.\ \ref{sec:premises}, in the GHC model, the linewidth is interpreted as a mixture of infall motions (often dominant) {\it and} turbulence, since the structures are assumed to be undergoing {\it turbulent (chaotic) gravitational contraction}. Therefore, the velocity dispersion $\sigma_v$, and the associated virial parameter $\av$ and Larson ratio $\Lcal$ are actually cloud {\it variables}, rather than parameters, which represent this mixture, {\it and evolve during the collapse}. As they evolve, they describe trajectories in the $\av$--$M$ and $\Lcal$--$\Sigma$ diagrams that coincide with the locus of observational determinations of these variables \citep[see eq.\ (8) and Fig.\ 2 in] [] {VS+19}.

\item The super-virial velocities often observed in low-column density objects in the KH diagram or low-mass clumps in the \aM\ diagram can be interpreted either as {\it assembly} by an external compression with a kinetic energy larger than that of the clump's self-gravity (either an external potential well, or a shock compression), or as {\it dispersal} \citep{Camacho+16, VS+19}. \citet{Camacho+16} found in numerical simulations that the proportion of assembling/disrupting clumps is roughly half and half. In neither case are the clumps in equilibrium, so there is no need for an equilibrating confining pressure.

\item The sub-virial velocities in high-mass or high column density clumps in the \aM\ and KH diagrams, respectively, are an indication that the structure is dominated by infall, but has not attained the full free-fall speed yet. This is because, on the one hand, for high-column density or high mass objects, the gravitational velocity is likely to be larger than any available turbulent velocity. On the other hand, because clumps begin to contract from a finite, rather than infinite radius, the infall speed is smaller than the free-fall speed, and only approaches it asymptotically \citep{VS+19}. The duration of the sub-virial stage is in fact expected to be long, because during the early stages, free-fall collapse proceeds very slowly \citep{Girichidis+14}, and moreover, the self-gravity is barely stronger than the thermal pressure \citep{Larson69}.

\end{enumerate}

\end{enumerate}

\end{enumerate}

\subsection{Lifetimes of structures and efficiency over a free-fall time} \label{sec:lifetimes}

On the basis of the inferred age spreads in young stellar clusters and associations, it has been argued \citep[e.g.,] [] {Tan+06, DaRio+14, Krumholz+19} that molecular clouds have lifetimes much longer (by roughly one order of magnitude) than their free-fall time $\tff$, and that, as a consequence, the so-called star formation efficiency per free-fall time \citep{Krum_McKee05}, 
\begin{equation}
    \epsff \equiv \frac{\dot M_*}{\Mg} \tff,
    \label{eq:eff}
\end{equation}
is small, of order 1\%.

\begin{enumerate}

\item {\it Interpretation in TS:} The clouds, and the filaments and clumps they contain, are {\it isolated entities} in a state of near-virial equilibrium, and therefore not collapsing, but rather, in a quasi-hydrostatic state \citep[e.g.,] [] {Larson81, McKee_Ostr07, Heyer+09, Field+11, Keto24}, globally supported by supersonic turbulence, albeit with strong local density fluctuations (the {\it cores}), caused by the turbulence-induced shocks, that can collapse locally \citep[e.g.] [] {VS+00, VS+03, Padoan+01, Padoan+20, MacL_Kles04, Ballesteros-Paredes+07}. Sometimes, the cores are assumed to also form in near equilibrium, and that they begin to undergo collapse only after the turbulence in their interior has dissipated \citep[e.g.,] [] {McKee_Tan03, Bergin_Tafalla07}, giving rise to {\it coherent} cores, with only mild, subsonic turbulence in their interior \citep{Goodman+98}. Since the structures are assumed to be in near equilibrium, they can have lifetimes much longer than their free-fall time. Also, because the only objects collapsing are the dense cores, the star formation involves a very small fraction of the total cloud's mass, therefore causing the observed small value of $\epsff$.

\bigskip
\item {\it Interpretation in GHC:} First of all, recent, detailed estimations of the age spreads revise them to values smaller by factors $\sim 2$ \citep{BonillaB+22}. Second, the clouds, and the clumps, filaments, and cores they contain, constitute a density hierarchy, with mass cascading from one level to the next, so that all density ranges accrete from their respective environments \citep[e.g.,] [] {Field+08, VS+09}, and their masses are continuously replenished by this accretion. Fundamentally, they are not ``objects'' consisting of roughly the same material throughout their lives, but rather they are {\it regions} in a continuous flow: the {\it loci} in the flow where gas transiting from low to high density takes a particular value of the density. In particular, they generally last longer than the {\it local} free-fall time (i.e., that associated to the region's density), because they are only a small part of the full collapse flow, which starts from a larger, lower-density structure with a longer free-fall time. Their lifetimes can then be of the order of the full duration of the accretion from the largest scales, perhaps approaching a stationary state \citep[e.g.,] [] {Murray_Chang15, Murray+17, Gomez+21, Naranjo+22}, if they are not destroyed first by feedback.

The mass $M_i(t)$ in density range $i$ (i.e., at the ``cloud'', ``clump'', ``filament'', ``core'', or ``fragment'' density level) is determined by the accretion rate {\it into} this level from lower densities, $\dot M_{{\rm in},i}$, and the loss rate onto the next higher level, $\dot M_{{\rm out},i}$. That is,
\begin{equation}
\frac{d M_i(t)} {dt} =  \dot M_{{\rm in},i} - \dot M_{{\rm out},i} 
\label{eq:mass_balance}
\end{equation}
The accretion onto the protostellar objects is only the last step of this cascade, but gravity plays a role in driving the accretion flow across {\it all} scales \citep{Burk_Hart13}. Importantly, recent simulations have shown that the gaseous structures' masses can continue to grow by accretion even while already forming stars \citep{GS_VS20}. This keeps the observed {\it instantaneous} efficiency, defined as 
\begin{equation}
{\rm SFE}(t) = \frac{M_*(t)} {M_{\rm g}(t)+M_*(t)},
\label{eq:SFE}
\end{equation}
at low values, because the gas mass $M_{\rm g}$ increases at an even faster rate than the stellar mass $M_*$.  \citet{VS+24} have proposed that the mass growth in the gaseous phase may originate from a {\it gravitational choking} mechanism. This consists in the inwards radial decrease of the accretion rate in a centrally-condensed gaseous region, if its spherically-averaged density profile is shallower than $r^{-2}$, when the accretion rate at every radius is driven by the gravity of the mass internal to that radius.

%{\color{red} LH I don't think this is necessarily true, as feedback can accomplish the same thing.} {\oli [EVS: We do observe this phenomenon even in simulations without feedback \citep{GS_VS20}, and so it must be operating at least to some level. Moreover, we don't see much of a difference in the growth rate of cores between the cases with and without feedback, so the role of the feedback is probably not dominant. I have mentioned this now in the text.]}

The inwards reduction in the accretion rate could also be attributed to the presence of feedback at various scales, starting with protostellar outflows. However, even in this case, numerical simulations show that the accretion flow onto the protostars is not halted \citep{Wang+10}. Moreover, the mass growth of the gaseous phase is observed even in simulations without feedback \citep{GS_VS20}, implying that the effect does not rely on some form of support.

In addition, observational determinations of $\epsff$ also come out low because they generally calculate the free-fall time for the {\it current} physical conditions, while the efficiency is calculated by counting the number of protostars and dividing by the total gas mass. Since the protostars have been forming for at least a few million years, their number corresponds to average collapse rates in the past, when the density was lower and the free-fall time was longer \citep{VS+19, BonillaB+22}.

The {\it final} SFE (the fraction of the total cloud mass that is turned into stars) remains low because the contraction occurs differentially, with higher-density gas contracting faster. Therefore, the first massive stars that form disrupt the rest of the cloud within a few to several Myr \citep[e.g.,] [] {Dale+12, Colin+13, ZA+19, Chevance+22, Grudic+22}. The accretion flow is also likely to be disrupted \citep[e.g.,] [] {Ginsburg+16, VS+24}, preventing the rest of the cloud's mass from reaching the star-forming, densest regions.

\end{enumerate}

\subsection{Cluster structure and dynamics}

Recent observations of young stellar clusters and associations using GAIA and APOGEE data \citep[see the review by] [and references therein] {Krumholz+19}, as well as combinations of X-ray and infrared observations in the MYStIX and SFiNCS surveys \citep{Feigelson+13, Getman+17}, have revealed a wealth of structural and kinematic properties of these stellar structures. Some important features of young stellar clusters are

\begin{itemize}

\item {\it Radial age gradients}. The stars located in denser parts of the clouds are younger and have smaller age spreads than stars located in less dense regions \citep[e.g.,] [] {Getman+14a, Getman+14b, Getman+18a, Getman+18b, Kounkel+18}.

\item {\it Extended and accelerating SF histories}. The age distribution in young clusters generally exhibits a tail of older stars, up to several-Myr old, a maximum at a more recent age, and often a drop at the youngest ages \citep[e.g.,] [] {Palla_Stahler99, Palla_Stahler00, DaRio+14, Lee+15, Caldwell_Chang18}. The extent of the SF activity is typically $\Dtsf \lesssim 10 \tffloc$, where $\tffloc$ is the {\it local} free-fall time in the star-forming region. However, $\sim 50\%$ of the stars have formed more recently, within $\sim 3 \tffloc$ \citep{Krumholz+19}.

\item {\it Moderate, chaotic expansion or contraction}. Young clusters generally exhibit from moderate \citep[$\lesssim 1 \kms$; e.g.,] [] {Kounkel+18, Kuhn+19} to non-discernible \citep{Ward_Kruijssen18} expansion or contraction, and with clear evidence for a scattered, rather than concentrated origin of the various subgroups conforming the cluster \citep[e.g.,] [] {Ward_Kruijssen18, Getman+18b}.

\item Massive star-forming regions (protoclusters) in which the spatial distribution of young stellar objects (YSOs) is more spatially scattered exhibit a shortage of the most massive stars that would be expected from direct calculation of the massive-star fraction expected from a standard IMF to the total number of existing protostars in the region \citep[e.g.,] [] {Povich+16, NguyenL+20}.

\begin{enumerate}

\item Interpretation in TS:
  \begin{enumerate}
    \item The long extent of the SF activity ($\sim 10 \tffloc$) in the cluster-forming clumps is attributed to a near-virial equilibrium between turbulence and gravity, with the only structures collapsing being the dense cores, formed by the external supersonic turbulence \citep[e.g.,] [] {Pad_Nord02, VS+03, McKee_Tan03, Krum_McKee05}. 
    
    \item The accelerating SFR in star-forming regions is attributed to the presence of a {\it conveyor belt-like} accretion flow onto the clouds, which causes their gas mass, and therefore the SFR, to increase over time, although the driver for this accretion flow is not specified \citep[] [Sec.\ 3.4.2] {Krumholz+19}. \citet{Padoan+20} attribute the accretion onto filaments, and from filaments to hubs, to {\it inertial inflows}, driven by large-scale turbulent compressions.
    
    \item To our knowledge, no explanation is provided by TS to the radial age gradients nor to the observed shortage of the most massive stars in regions where the star formation activity is more spatially scattered.
    
  \end{enumerate}
  
  \item Interpretation in GHC:
  \begin{enumerate}
    \item The long extent of the SF activity ($\sim 10 \tff$) in the cluster-forming clumps is attributed to an ongoing gravitational contraction and the continuous accretion it drives onto all scales of the hierarchy of structures in MCs. As a consequence, the relevant free-fall time for the accumulated stellar content in the clumps is that of the {\it initial} condition, which had a significant lower density, and therefore a significantly longer $\tff$ \citep{VS+19}.
    
    \item Similarly to the interpretation in TS, the accelerating SFR in star-forming regions is explained as a consequence of conveyor belt-like accretion flows that cause the gas mass, the mean density of the clumps, and therefore the SFR, to increase over time \citep{GV14}. However, contrary to the case of TS, this accretion is in general attributed to the gravitational contraction of the cloud or clump's  parent structure, as gravity is observed to be important at all scales in numerical simulations covering the required range of scales \citep[] [see also Sec.\ \ref{sec:global} above] {GV14, Camacho+23}.

    \item The initial radial age gradients are interpreted in GHC as a consequence of the faster accretion onto the central hubs than onto the secondary star-forming sites along the filaments. This causes the central hub to become massive faster than the secondary sites, making it the main star-forming site at advanced stages in the complex \citep{VS+24}. Therefore, at these advanced stages, most of the stars form at the hub, causing the young stars to be concentrated there. However, since it takes time for even the central hub to become massive, at early times the difference in the SFR between the hub and the peripheral sites was not so large, and so the older stars are more scattered than the younger ones \citep[see the evolution of the stellar production in Fig.\ 2 of][]{GS_VS20}. 
    %This process is analogous to the ``competitive accretion'' scenario \citep{Bonnell+01}, but at the hub-core level, rather than at the individual protostellar level \citep{VS+19}.

    \item The observed scarcity of massive stars in spatially-scattered regions is attributed to the fact that the development of a dominant, central hub capable of forming massive stars requires relatively long times, typically of a few to several Myr, necessary for the hub to accrete enough mass from its surroundings \citep{VS+24}. Regions lacking the required concentration cannot yet form significantly massive stars, even if there are numerous low-mass star-forming sites.
    
  \end{enumerate}

\end{enumerate}

\end{itemize}

\subsection{The relative orientation between dense molecular cloud filaments and the magnetic field} \label{sec:mag_field_fils}

The magnetic field in the neighborhood of dense {\it Herschel} molecular cloud filaments is observed to be preferentially perpendicular to them, although simultaneously parallel to striations that appear to accrete onto the filaments \citep[e.g.,] [] {Palmeirim+13}. \citet{Soler+17} presented one of the now most accepted interpretations to this observation, which remains agnostic as to whether turbulence or gravity dominate the dynamics. They derived an evolution equation for the angle $\phi$ between the density gradient and the magnetic field vector, and showed that the configurations where the field and the density gradient are either parallel or perpendicular are equilibrium points of the evolution. From this result, they suggested that the system may tend to evolve to either of them; i.e., that these configurations {\it may be} attractors.\footnote{Note, however, that equilibrium states are not necessarily attractors. For this to occur it is necessary that the equilibria are {\it stable}. \citet{Soler+17} did not investigate the stability of these equilibria.} In addition, they further found that the elongated density structures and the field tend to be aligned in low density regions where the velocity gradient tensor is dominated by shear, while they tend to be perpendicular at high densities, where the velocity gradient is dominated by compression.

\begin{enumerate}

\item {\it Interpretation in TS.} To our knowledge, no causal explanation is provided in the TS scenario, and the interpretation of this phenomenon in this scenario simply adopts the generic proposal of \citet{Soler+17}.

\bigskip
\item {\it Interpretation in GHC.} On the other hand, within the GHC scenario, a more direct cause-effect interpretation invoking gravity was provided by \citet{Gomez+18}. These authors noted that, if the energy in the multi-stage, anisotropic, gravitationally-driven accretion flow from cloud to filament and from filament to hub dominates over the magnetic energy, then the flow orients the field by dragging and stretching it in the flow direction. In this case, the cloud-to-filament flow, which is mostly perpendicular to the filament, causes the field in the neighborhood of the filament to be mostly perpendicular to it. However, as the accretion flow approaches the filament, it smoothly changes direction, and continues along the filament onto the hub. As a consequence, the field must also show a tendency to become parallel to the filament inside it. However, \citet{Gomez+18} showed that diffusive processes prevent the field from becoming fully parallel to the filament, and instead it is expected to develop a ``U'' shape across the filament. Several observational studies have found observational evidence of this process \citep{Pillai+20, Wang+20, Wang+24, Arzoumanian+21, Ching+22, Rawat+24}.

Note that the interpretations of \citet{Soler+17} and \citet{Gomez+18} need not be mutually exclusive. Instead, it is likely that they are complementary, with gravity providing the predominantly compressive nature of the velocity gradient in the self-gravitating regions and other processes, such as the nonlinear thin-shell instability \citep{Vishniac94}, providing the shear in non-self-gravitating regions (Granda et al., in prep.). However, it should be remarked that, within the TS scenario, no analytic model nor numerical simulation has shown that the required smooth changes of direction of the accretion flow can be produced without the action of gravity.

\end{enumerate}

\section{Similarities, differences, and transitional features} \label{sec:sim_diff}

\subsection{Both involve turbulence} 

In TS, the turbulence is consistently {\it strongly} supersonic, with sonic Mach numbers $\Ms \sim 5$-20 at the larger scales, and maintains the clouds and their substructures in near virial equilibrium. Importantly, {\it it is assumed that the nonthermal velocity dispersion derived from molecular lines corresponds exclusively to turbulence}, throughout the entire hierarchy of structures, from clouds to cores. It is also capable of inducing the collapse of the small-scale density fluctuations it produces \citep[e.g.,] [] {VS+03, MacL_Kles04}.

%    {\color{red} JBP: it is important in this paragraph to make the distinction between turbulence and linewidth. La turbulencia puede ser transónica, pero el ancho de línea bien puede ser supersónico porque se tienen movimientos hacia diferentes centros de colapso con velocidades relativas mayores a 0.2kms. } 
In GHC, on the other hand, the internal turbulence generated by various fluid instabilities during the cloud's assembly process is only {\it moderately} supersonic, and is {\it not} able to balance gravity nor to induce local collapse \citep{VS+07, GG_VS20, Zavala-Molina+23}. The strongly supersonic, yet near-virial, motions observed in molecular clouds are {\it a result} of the gravitational contraction before stellar feedback becomes dominant \citep[e.g.,] [] {VS+07}, and generally reflects the infall motions onto multiple, hierarchical collapse centers, and their relative motions (see Sec.\ \ref{sec:GHC_def}, item \ref{sec:item_non-monolithic}).
    %{\color{red} LH you mean the "turbulence" generated by non-gravitational forces.} {\oli EVS: I have rephrased this paragraph (and added the next one) to more accurately describe the various sources of turbulence and what each can do.}
This moderately supersonic turbulence only generates moderate density fluctuations that act as seeds for the later collapse of the small-scale structures \citep{Clark_Bonnell05}, once the global Jeans mass has decreased by global contraction to the level of those structures' mass \citep{VS+19}. {\it The linewidths are interpreted as a mixture of infall and turbulent motions}, with the truly turbulent motions accounting for only $\sim$ 30--40\% of the linewidth, and being generated by the chaotic nature of the multi-center, hierarchical gravitational collapse \citep{GG_VS20}. Finally, after feedback becomes dominant, it drives strongly supersonic {\it and} super-virial motions that disperse the clouds and causes $\av \gg 1$ \citep{Colin+13}.

Additionally, it is important to acknowledge that the turbulence injected by stellar sources can have various effects. Photoionization, winds, and Type II-supernova feedback from massive stars inside the clouds can rapidly disperse them \citep[e.g.,] [] {Colin+13, Haid+19}, while external Type Ia SN explosions can either provide the converging flows assembling the clouds or partially disperse them \citep[e.g.,] [] {Iffrig_Henneb15}.

It is important to note that {\it neither} model requires turbulence to be dissipated inside cores, to produce the so-called ``coherent cores''. Since the turbulent cascade implies that the velocity difference between two points decreases with decreasing separation between them, and so subsonic structures necessarily exist below some scale called the {\it sonic scale} \citep[e.g.,] [] {VS+03, Klessen+05, Federrath+21} without implying that the turbulence dissipation scale has been reached. In fact, the turbulence at scales smaller than the sonic scale may be fully developed, yet subsonic. The dissipation scale of the turbulent motions may be much smaller than the molecular cloud core size.

\subsection{Both are hierarchical} 

In TS, the hierarchy is produced by the scale-free nature of turbulence, and consists of eddies within eddies, with the notion of ``eddie'' being generalized to include ``shocklets'' (turbulent density fluctuations) in the supersonic case. 
    
%{\color{red} JBP: it is important to acknowledge that some hierarchy should be inherited from the large-scale process that formed the cloud in first place.} 
In GHC, the hierarchy is primarily caused by sequential gravitational fragmentation due to the continuous reduction of the Jeans mass during the collapse, and therefore consists of collapses within collapses, although some of the structuring (e.g., large-scale filamentation) may be due to the processes forming the clouds themselves (inertial compressions, and the thermal, nonlinear-thin-shell and Kelvin-Helmholtz instabilities). The gravitational fragmentation is {\it seeded} by the moderate turbulent density fluctuations, whose hierarchical nature is also recognized, so that there exist turbulent density fluctuations at all scales. The collapse of smaller-scale structures occurs later in time, because it can only start after the global Jeans mass in the cloud has decreased (by the global contraction) to the value of the small-scale structures' mass \citep{VS+19}.
 
\subsection{Both imply long lifetimes of the dense structures} 
  
In the early versions of TS, the long lifetimes were assumed to be caused by turbulent support at all scales, except at the core scale ($\sim 0.1$ pc), where gravity is assumed to take over and induce collapse \citep{VS+03, MacL_Kles04, Krum_McKee05}. In more recent years, it has been realized that, except at low masses or column densities, large values of the virial parameter are observed in the $\av$-$M$, and $\Lcal$-$\Sigma$ diagrams, and thus turbulence is assumed to dominate. These objects are assumed to be either pressure-confined \citep[e.g.,] [] {Field+11} or expanding. 

In GHC, since dense structures are flow features, lifetimes longer than the {\it local} free-fall time are caused by continuous replenishment via accretion from the next larger scale. Nevertheless, they are not assumed to be as long ($\sim 10$ free-fall times) as in the TS scenario, in accordance to recent observational estimations \citep[e.g.,] [] {BonillaB+22, KimJ_Chevance+23}

\subsection{Both imply chaotic motions} 
  
In TS, the chaotic motions are the manifestation of the turbulence. In GHC, they are a mixture of moderately turbulent motions and predominantly gravity-driven infall motions towards multiple collapse ``centers'', creating an intricate and hierarchical multi-focus accretion flow. The accretion centers need not be point-like, but can also be planar or filamentary. This is the manifestation of the amplification of anisotropies by pressureless collapse \citep{Lin+65}. 

\subsection{Both imply gravitational domination and contraction in some range of scales}

In the original versions of TS,\footnote{As further discussed in Sec.\ \ref{sec:model_convergence}, this hypothesis has been relaxed to some extent over the years within the context of TS, to acknowledge the collapse up to the clump ($\sim$ pc) scale, thus allowing for {\it clump-fed accretion} \citep[e.g.,] [] {Wang+10}.} only the smallest scales ($\lesssim 0.1$ pc), below the {\it sonic scale}, for which the turbulence is subsonic, are assumed to collapse freely, if they are also Jeans-unstable \citep[e.g.,] [] {VS+03, MacL_Kles04, Krum_McKee05, Ballesteros-Paredes+07,  McKee_Ostr07, Bergin_Tafalla07}. At larger scales, the collapse criterion is the ``turbulent Jeans mass'', in which the sound speed is replaced by the turbulent velocity dispersion, and the density may be assumed to be mean density multiplied by the square of the turbulent Mach number, as that is the typical density expected for the turbulent density fluctuations \citep[see, e.g., Sec.\ IV.E of] [eq.\ (2) of \citealt{Palau+15} and eq.\ \eqref{eq:turb_MJ_pract} in this paper] {MacL_Kles04}.

In GHC, clouds may be gravitationally dominated at all scales, until destroyed by the feedback from their internal stellar products. External feedback can perturb the outskirts of the clouds, but this driving (SN explosions) is impulsive (i.e., momentary) and temporally intermittent, in nonlinear dynamics terminolgy (i.e, sporadic), while gravity acts continuously, as seen in the animation of Fig.\ \ref{fig:straight-arm_sim}. The collapse criterion continues to be the thermal Jeans mass because the turbulence driven continuously by the collapse is subdominant \citep{GG_VS20}. Low-column density and/or low-mass objects \citep[in general, objects of weak self-gravity;] [] {Camacho+23} that are dominated by the kinetic and/or magnetic energies are often in the process of assembly \citep{Camacho+20} by external compressions that may be either inertial or driven by a larger-scale gravitational potential. Of course, these super-virial structures may also be truly dispersing \citep{Camacho+20, Ganguly+22}. 

\section{Existing evidence and further suggested tests} \label{sec:tests}

Since the two models interpret the same observational features of molecular clouds and their substructures in terms of different mechanisms, further predictions are needed in order to distinguish between the models. Some efforts have already been conducted in this direction \citep[e.g.,] [] {Getman+18a, Getman+19, Chen+19, Sanhueza+19, Beuther+20, Dewangan21, Chen+22, Smith+22, Ganguly+22}, although a precise definition of the expectations from each model is necessary in order to avoid confusion. In this section we list a number of additional suggested tests.

\subsection{What fraction of the linewidth corresponds to infall and what fraction corresponds to turbulence?} \label{sec:linewidths_turb_coll}

Both virial equilibrium and free-fall predict essentially the same energy-budget signatures, since in both cases $\Ek \sim |\Eg|$. 
%However, collapse occurs on short timescales, while quasi-equilibrium structures evolve on much longer ones. Under TS, the existence of stars significantly older than the {\it local} free-fall time has been interpreted as a signature that star-forming regions are supported by turbulence against collapse \citep[e.g.,] [] {Krumholz+19}. However, under GHC, this same property is interpreted as evidence of accretion replenishing the gas \citep{VS+09}, the latter nevertheless infalling at speeds comparable to the free-fall velocity.
%
Linewidths corresponding to speeds comparable to the free-fall speed are routinely observed in clouds, clumps and cores, as manifested in the fact that these objects have virial parameters that hover around unity \citep[e.g.,] [] {Heyer+09, BP+11, BP+18, MivilleD+17, Traficante+18b, Camacho+23}. This property has been generally interpreted as evidence of the objects being in near equilibrium \citep[e.g., ] [] {Larson81, Heyer+09, Field+11}. However, an important question then is whether the linewidths are really dominated by microscopic turbulent motions providing support or by large-scale infall motions. Some pioneering works attempting to resolve this ambiguity have provided evidence that the linewidths may correspond to actual infall motions, rather than to turbulence providing support \citep[e.g.,][]{Barnes+18, Traficante+18a, Traficante+18b, Traficante+20, Zhou+24b, Shimoikura+25}{}{}. 

In particular, \citet{Traficante+20} have suggested that the infall contribution dominates at surface densities $\Sigma\gtrsim 500 \Msun$ pc$^{-2}$, 
% 0.1 g cm^-2
at least at the parsec scale, while, examining velocity gradients in  the nearby galaxy NGC628, \citet{Zhou+24b} have shown that entire molecular clouds act as hubs of hub-filament systems, accreting from their surrounding, lower density molecular gas. Similarly, \citet{Shimoikura+25} have shown that chemically young molecular gas in the Polaris molecular cloud exhibits velocity gradients indicative of accretion from the surrounding, lower-density molecular gas, suggesting gravitational domination at column densities as low as $\sim 10^{21} \psc$.  In any case, further studies along these lines and on different classes of objects are necessary. Promising tests to separate microturbulence from other systematic motions could be based on the velocity structure function \citep[e.g.,] [] {Henshaw+20, Palau+21}, or on comparing the projected velocity gradient with the 2D gravitational potential gradient obtained from column density maps, to approximately determine the fraction of the kinetic energy in the velocity component parallel to the gravitational acceleration vector.

\subsection {Is the Larson ratio $\Lcal$ constant or does it scale as $\Sigma^{1/2}$?} 

If the velocity dispersion in clouds indeed only represents turbulence, and the velocity dispersion-size Larson scaling relation has indeed its origin in the Burgers-like turbulent energy spectrum ($\Ek \propto k^{-2}$) expected for strongly supersonic turbulence \citep[e.g.,] [] {VS+00, MacL_Kles04, Elm_Scalo04, Kritsuk+07} as maintained in TS, then $\Lcal$ should be the same for all regions in which the turbulence driving source is the same. Conversely, if the velocity dispersion is dominated by gravitational contraction motions, with only a sub-dominant fraction of truly turbulent motions (except during the early assembly stages), as proposed in GHC, then the Keto-Heyer scaling $\Lcal \propto \Sigma^{1/2}$ is expected, except in the objects with the lowest $\Eg$ in each mass or density range.

In fact, this test has already been performed by numerous observational \citep[e.g.,] [] {BP+11, BP+18, Leroy+15, MivilleD+17, Traficante+18a} and numerical studies \citep{Camacho+16, Camacho+23, IbanezM+16, Padoan+16}. In particular, the latter two studies have shown that the $\Lcal$-$\Sigma$ scaling is {\it not} satisfied in simulations without self gravity or with very strong turbulence driving. In any case, more systematic tests need to be carried out, in order to confirm or reject the hierarchical nature of the scaling predicted by GHC, as it should be repeated at each stage of the gravitational fragmentation hierarchy.

In addition, \citet{Peretto+23} have reported measurements of the virial parameter at different scales for a sample of 27 IRDCs, finding that the clouds seem to be self-gravitating at all scales, but that parsec-scale clumps exhibit steeper density profiles ($\rho \sim r^{-2}$) than those of their enveloping clouds, and nearly flat velocity dispersion profiles ($\sigma \sim r^0$). They interpret these results as implying that the clumps are collapsing, while their enveloping clouds are stable, although it is also possible to interpret their results as indicative of a different mode of collapse of the enveloping gas, if, for example, having a lower density, it is evolving on longer timescale, and therefore it is still in  transient stage in which the density profile is steepening in time, as discussed by \citet{VS+24}. 

%{\cyan AP: Recently, \citet{Peretto+23} report }

%\subsection {\it How does $\av$ scale with region size for nested structures?} As discussed in Sec.\ \ref{sec:Larson_rels}, Larson's linewidth-size relation can only be simultaneously consistent with both supersonic turbulence {\it and} virial equilibrium for objects that satisfy Larson's density-size relation (i..e., that have roughly the same column density). Once objects of significantly different column densities are considered, the Larson linewidth-size relation breaks down \citep[e.g.,] [] {Caselli_Myers95, Plume+97, Shirley+03, Gibson+09, Wu+10}, and instead the Keto-Heyer relation, eq.\ \eqref{eq:Larson_ratio} is satisfied. In order for structures of {\it all} column densitite sto be balanced by turbulence, it would be necessary that the turbulence at different column density ranges belonged to different cascades. Otherwise, denser structures would appear sub-virial.

%Alphavir should be smaller at larger scales. This could be related to the well-known alphavir vs mass relation. 

\subsection{Is the excess kinetic energy in supervirial clumps driven by a larger-scale gravitational potential?} \label{sec:excess_Ek}

As discussed in Sec.\ \ref{sec:devs_vir}, low-mass objects in the \aM\ diagram, and low-column density objects in the \LS\ diagram exhibit a large scatter, ranging from being nearly-virial to strongly supervirial at a given mass or column density, respectively \citep[e.g.,] [] {Keto_Myers86, Kauffmann+13, Leroy+15, MivilleD+17, Traficante+18b}. As also discussed in that section (Sec.\ \ref{sec:devs_vir}.4), under the TS scenario, the supervirial nature of {\it small} objects is understood as a consequence of the Larson linewidth-size relation and a constant-density assumption, which together imply $\av \propto R^{-1}$. Instead, under the GHC scenario, the supervirial nature of {\it weakly gravitationally bound objects} (regardless of size) is interpreted as a manifestation of the evolution of all objects, which necessarily starts with an external (i.e., not due to the objects' self gravity) compression, which in turn increases the objects' self gravity, and so the compression transits from being externally driven to driven by self-gravity. 

However, recent studies from \citet{Gomez+21}, \citet{Ganguly+22} and \citet{Camacho+23} have shown that often the external compressions are due to the gravitational collapse of the objects' parent structures. In this case, gravity is still the dominant driver, but it is the self gravity of the parent structure, within which the weakly-bound object is just the ``tip of the iceberg'', being compressed by the ram pressure of the infall of the outer regions onto it. 

Therefore, an important test is to determine whether the 
gravitational binding of {\it hierarchically nested} structures increases (as expected in GHC) or decreases (as expected in TS) outwards, both in observations and in numerical simulations.

%{\color{red} LH It seems to me that an entire paper could be written about the difficulty of calculating/defining a (the) virial parameter. Observatonally, you typically only have one velocity dimension, so one must assume isotropy, which is not very likely. But the worst problem is the gravitational potential in a medium with very large density contrasts. That's true even for simulations (See Figure 14 in Kuznetsova et al. 2015) and is even harder to figure out observationally.
%}

%{\oli EVS: I agree that calculating $\av$ is tricky and that there's a lot of ambiguities involved. However, from a practical point of view, people do measure it and make a lot of inferences from their measurements, so I'd say we should try to understand what is being measured, and then inform the community about its meaning. Also, we could provide some recommendations on the best ways of making the measurements. We certainly will not convince the community that $\av$ is useless.}

\subsection{Is the magnetic field strong enough to support observed subvirial massive cores?} 

Several recent observational studies have reported the existence of massive cores that are significantly {\it subvirial} \citep[e.g.,] [although see \citealt{Singh+21} for an opposite view] {Tan+13, Kauffmann+13, Pattle+17, Barnes+21}, and proposed that, in order to be virialized, as in the TS interpretation, they would need to be supported by the magnetic field. Instead, under GHC, they may be in the early stages of gravitational contraction, in which the infall speed has not yet reached the free-fall value, because the collapse did not start from infinity, but rather from a finite radius \citep{VS+19}, and furthermore, the collapse may be slowed down by the magnetic field even in collapsing super-critical cores. Therefore, measurements of the field strength and of the infall speed are necessary to discriminate between the two scenarios in these objects.

Another common interpretation \citep[e.g.,] [] {Liu+15} is that subvirial massive cores may be on the ``verge'' of gravitational collapse. However, in order for this to actually happen, it would be necessary that the objects have a mass of the order of the thermal Jeans mass, since otherwise there would be no support at all, and they should be undergoing collapse already. Under GHC, the interpretation is instead that they are already collapsing, although they have not reached the free-fall speed yet because the collapse has started only recently \citep{VS+19}. A relevant test in this case could be to estimate the number of Jeans masses that the cores contain.
%In this case, an estimate of the density contrast with respect to their background could provide an estimate of the time since the onset of collapse, which can then be compared to the background free-fall time, and thus estimate the expected infall speed at that stage of the collapse.} 

%{\color{red} LH I think it is essential to point out that these papers refer to {\em massive} cores, which are very different objects than low-mass cores.}
%{\oli EVS: Done.}

\subsection {How does the mass-to-flux ratio scale with region size around local star-forming sites?} 

According to GHC, the mass-to-magnetic flux ratio should be larger at larger scales {\it in the ideal-MHD case} \citep[i.e., if no strong diffusive mechanism such as ambipolar or reconnection diffusion is already dominating the magnetic field evolution;] [] {VS+05, Gomez+21}. However, if diffusive mechanisms are operating, then the mass-to-flux ratio is indeed expected to increase inwards.

Recent works by \citet{Hwang+21}, \citet{Koley+22}, \citet{Yen+23} and \citet{Liu+24_masstofluxratio} suggest a picture where the mass-to-flux ratio increases closer to the star-forming sites, and that a subcritical envelope surrounds a supercritical core, in line with the early work of \citet{Koch+12}. These results suggest that diffusive effects are dominating in the central cores. However, other recent studies reveal the opposite trend for the mass-to-flux ratio \citep[and the virial parameter, e.g.,][]{Tang+19}. In addition, the multi-scale study of \citet{Koch+22} in the W51 complex suggests collapse from pc-scales overcoming the magnetic field. The current observations in this direction are scarce and contradictory, and a systematic study in a large sample is required to establish more solid conclusions, especially including a determination of whether a diffusive process dominates or not.
%Liu, Zhang, Girart+24 show in their Fig. 17a clearly a trend of decreasing mass-to-flux ratio w radius, the opposite to what was predicted in Gomez+21.
%EVS answer: Una respuesta rápida es que nuestra predicción de que M/phi disminuye hacia el centro sólo es válida mientras se mantenga la condición de MHD ideal. En cuanto empieza a ser importante cualquier mecanismo difusivo (difusión ambipolar o difusión por reconexión), es de esperar que M/phi vuelva a aumentar hacia densidades mayores. Entonces, hay que ver en qué régimen están y a qué escalas de tamaño. Es un punto que ciertamente tenemos que dejar claro en el TS vs GHC. Es decir, nuestro "claim" no sustituye a las viejas ideas de difusión ambipolar, sino que sólo las inserta dentro de un contexto más amplio de cómo se comporta M/phi con el radio en general. 
 
%\subsection{What is the structure and magnitude of the velocity field in and around the dense structures (filaments and cores)?} 

\subsection{The correlation between the velocity and density fields}

\subsubsection{Spatial correlation between velocity dispersion and column density} \label{sec:sigma_N_correl}

Because the origin of the majority of molecular gas structures (MCs and their substructures, clumps, filaments, and cores) is different in the TS and GHC scenarios, different correlations between the velocity and density fields are expected in each model. In TS, the structures from MC down to core scales are assumed to be formed by strong shocks \citep{Padoan+01}, and so the dense structures constitute essentially {\it post-shock} regions---the downstream flow from a shock. In consequence, under TS, one should expect the density maxima to be offset from the shock fronts. Observationally, if the shocks correspond to local velocity dispersion maxima, or to emission in specific tracers such as the SiO molecule, under the TS scenario one should expect the velocity dispersion (or shock tracer emission) maxima to be {\it offset} from the column density maxima \citep[e.g.,] [] {Klessen+05}. Along these lines, \citet{Abe+23} have recently modeled the stabilization by ambipolar diffusion of a slow MHD shock bounding a filament. 

Conversely, under the GHC scenario, the majority of the structures are expected to be formed by gravitational contraction, which, away from the local centers of collapse, produces relatively smooth flows (see, e.g., GV14), similar to the case for Bondi accretion flow. Strong collisions, and thus strong turbulence are only produced at the local collapse centers (the core and hub centers), where the gas streams infalling through the various filaments collide with each other. The resulting strong turbulence causes a large velocity dispersion. Therefore, under GHC, the velocity dispersion (or shock tracer) maxima are expected to be nearly coincident with the column density maxima.

\begin{figure*}
\centering
\includegraphics[width=0.45\linewidth]{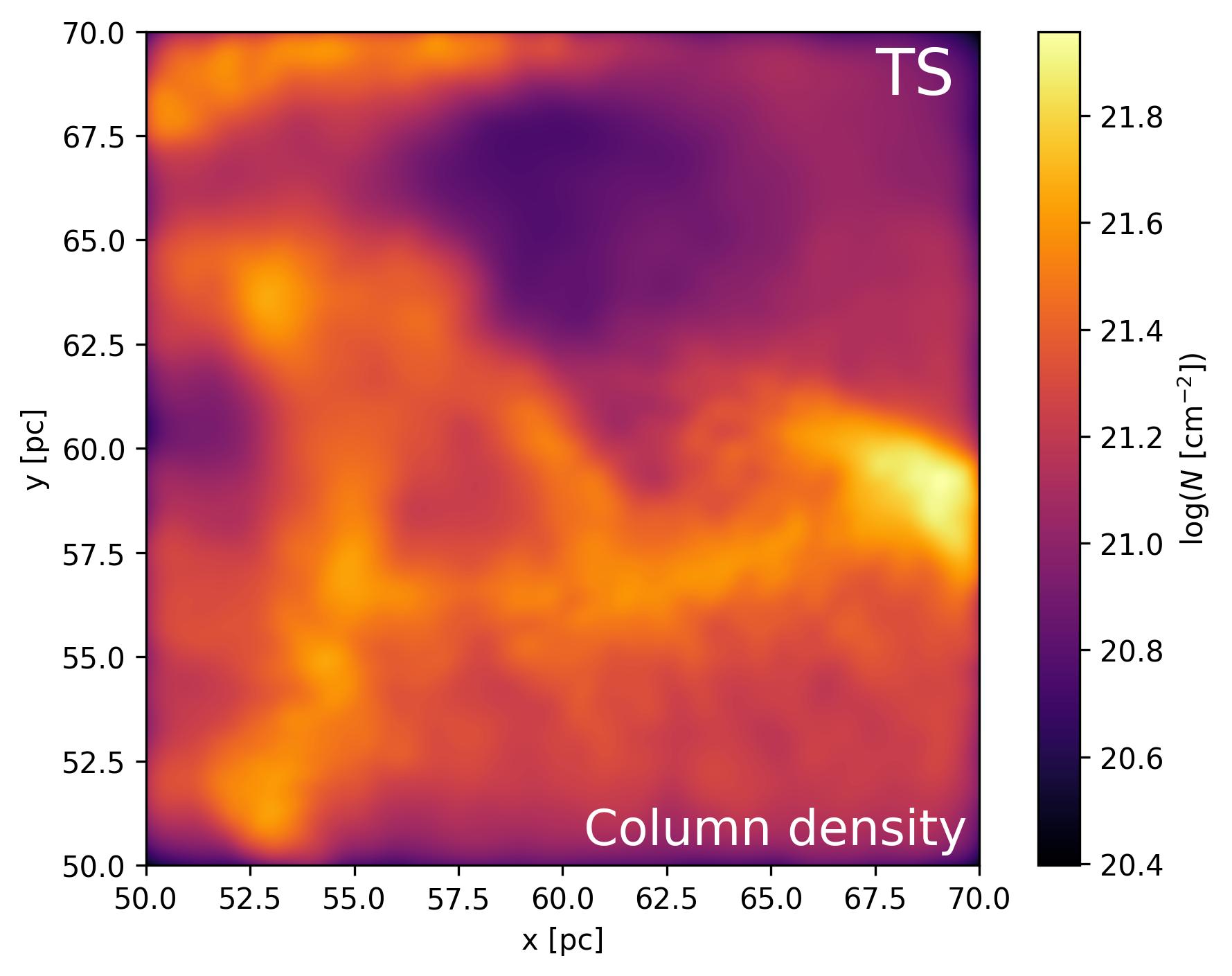}
\includegraphics[width=0.45\linewidth]{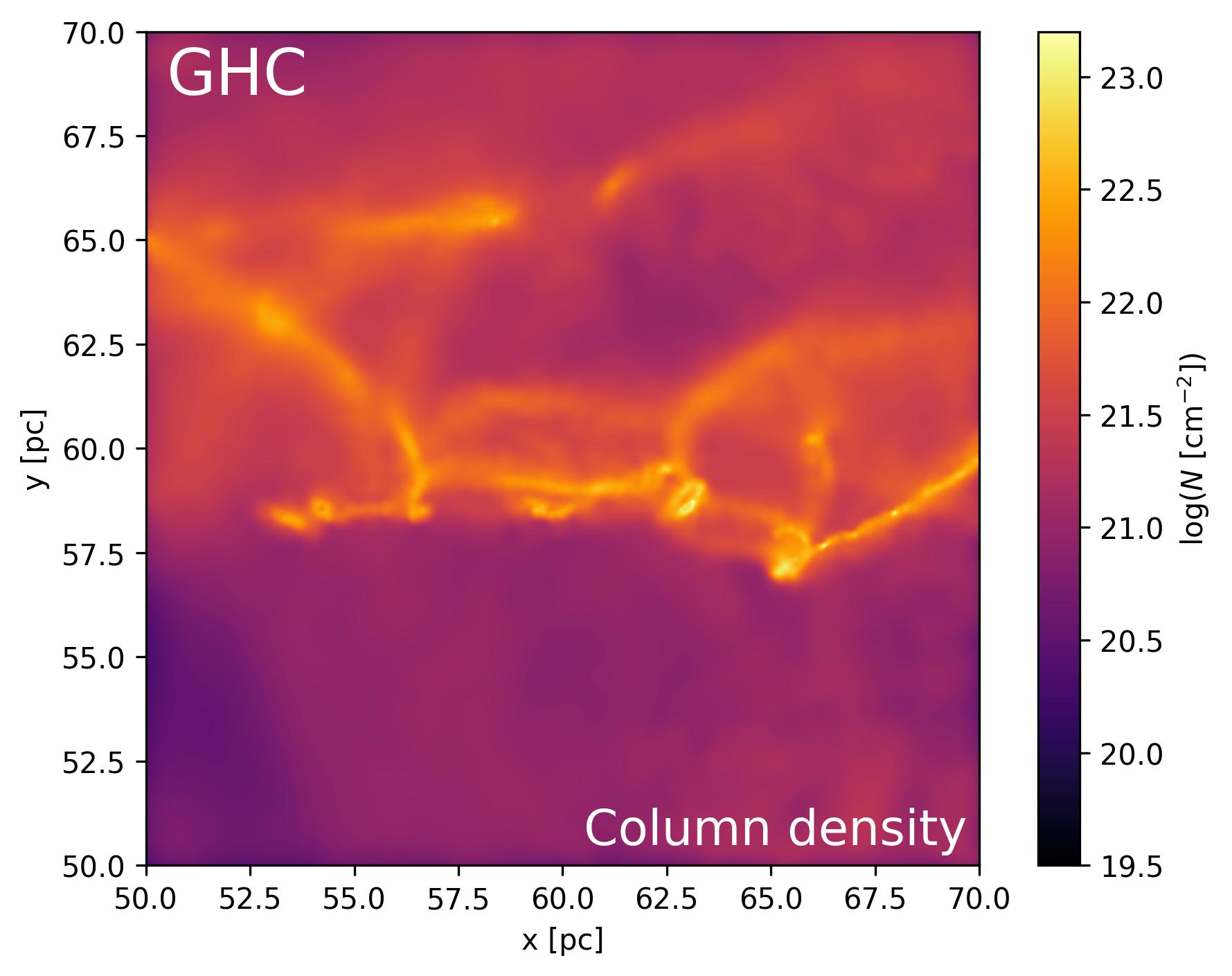}
\includegraphics[width=0.45\linewidth]{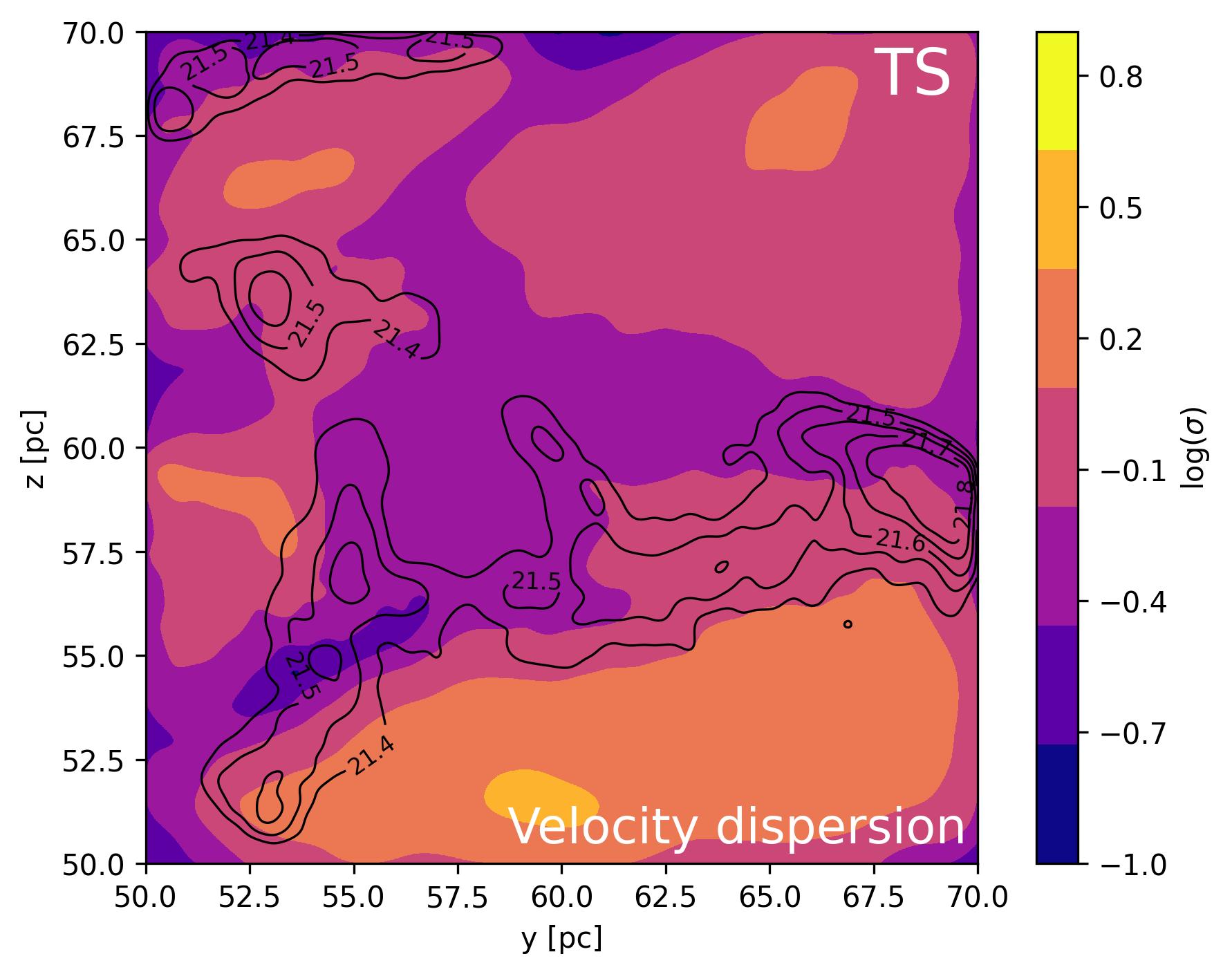}
\includegraphics[width=0.45\linewidth]{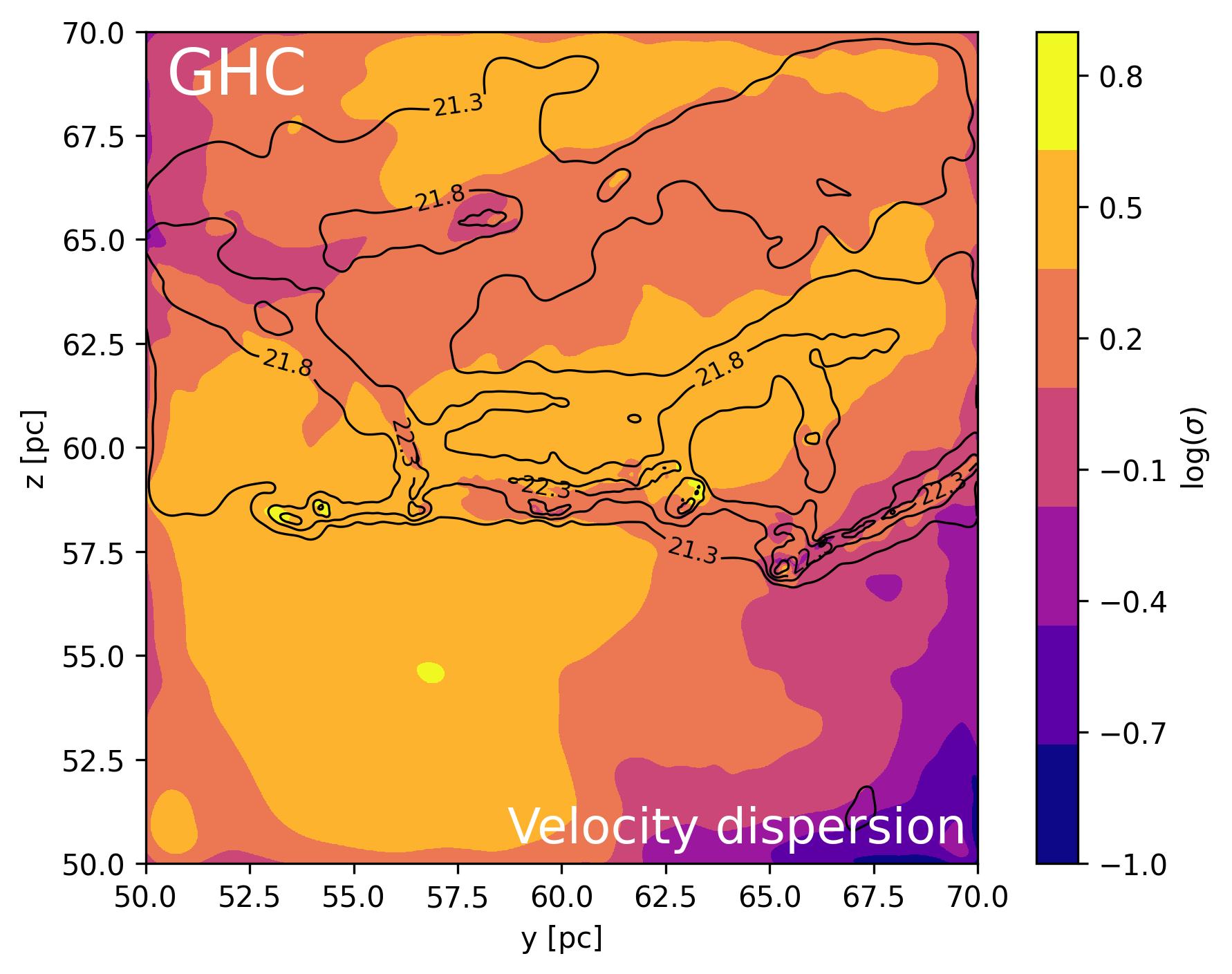}
\includegraphics[width=0.45\linewidth]{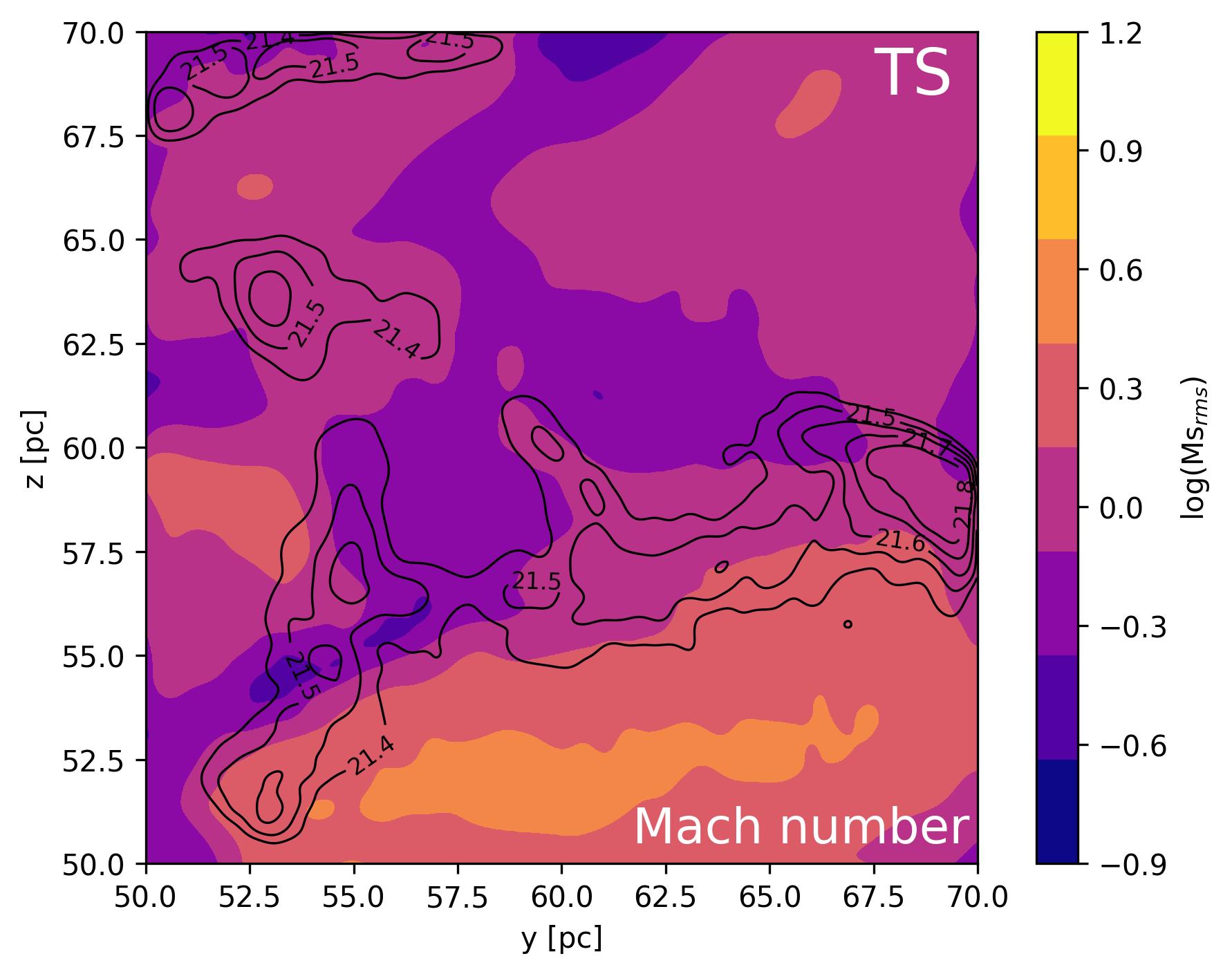}
\includegraphics[width=0.45\linewidth]{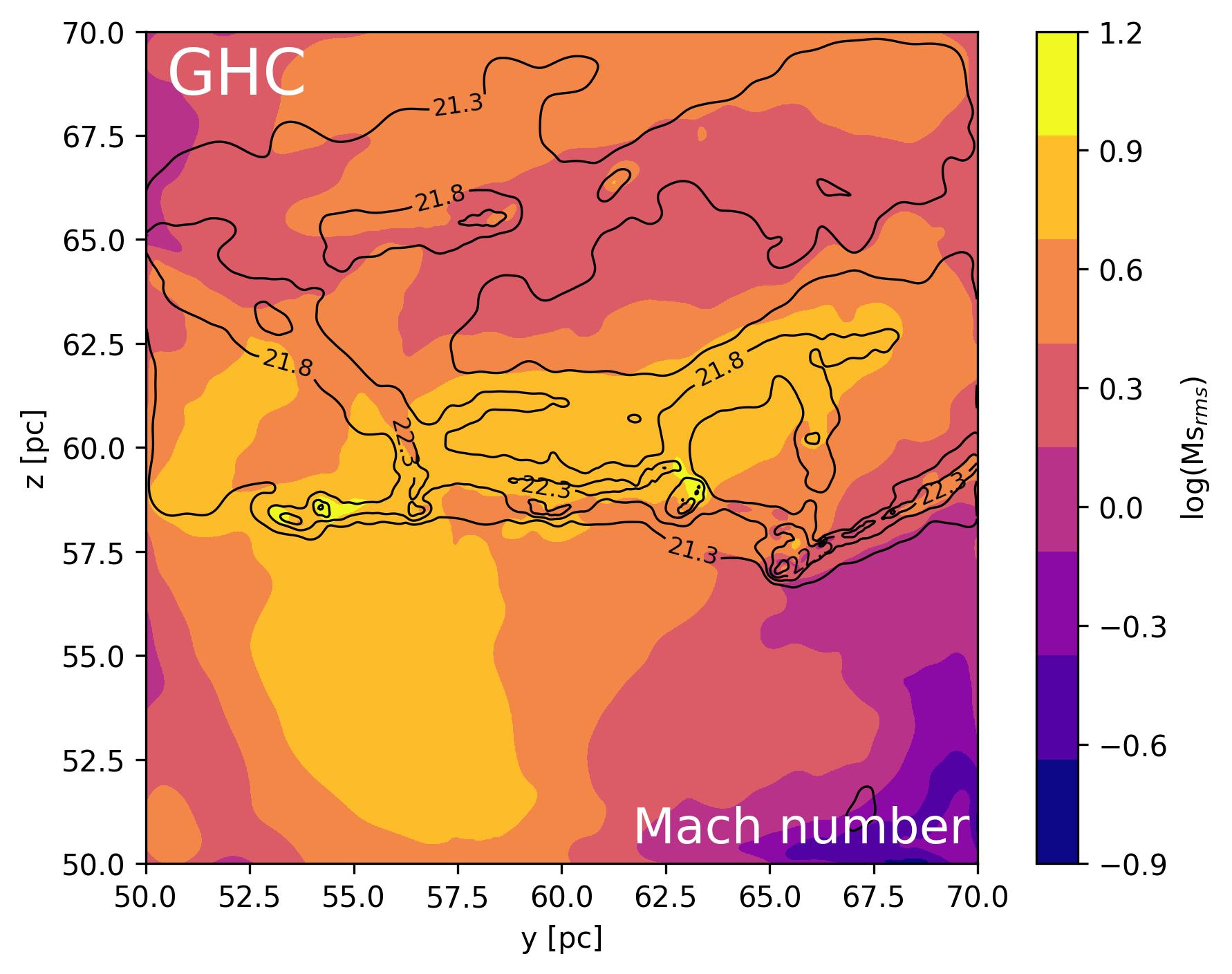}
\caption{Projected images of the column density ({\it top panels}), velocity dispersion ({\it middle panels}) and sonic Mach number ({\it bottom panels}) of a 20-pc-per side region within 256-pc simulations of decaying turbulence in the WNM, one without self-gravity ({\it left panels}) and one with it ({\it right panels}). The middle and bottom images have overlayed contours of column density. It can be seen that the velocity dispersion and column density maxima are offset from each other in the simulation without self gravity ({\it bottom left panel}), but are coincident in the simulation with self gravity.}
\label{fig:sigma_N_maps}
\end{figure*}

We illustrate this effect in Fig.\ \ref{fig:sigma_N_maps}, where we show maps of the column density ({\it top panels}), velocity dispersion ({\it middle panels}) and sonic Mach number of the same region in two simulations of decaying-turbulence in the WNM, one without self gravity ({\it left panels}) and the other with it ({\it right panels}; see Appendix \ref{app:phantom_simulation} for a description of these simulations), but otherwise identical. Note that the simulations include cooling representing the multiphase atomic gas. It can be clearly seen from these images that the velocity dispersion and Mach number maxima tend to coincide with the column density peaks in the presence of self gravity, while they tend to be located {\it away} from the density peaks in its absence. A detailed study of the spatial correlation between projected velocity dispersion and column density will be presented elsewhere.

\subsubsection{Comparison to observations}

The predictions from both models, can be discussed in the light of existing observations. First, we note that the tendency of the velocity dispersion (or Mach number) to increase together with the column density observed in the simulation with self gravity (Fig.\ \ref{fig:sigma_N_maps}, right panels) is qualitatively consistent with the trend seen in observational data \citep[e.g.,] [] {Busquet+13,Barnes+18, Suri+19, Beuther+20, Zhou+24a, GutierrezChaves+24}. Furthermore, this trend appears consistent with the result found by \citet{Arzoumanian+13} that more massive, supercritical filamentes tend to have larger, supersonic velocity dispersions. Instead, the offset seen in the purely turbulent simulation without self gravity (left panels of Fig.\ \ref{fig:sigma_N_maps}) is in general not seen in observational data, although it would be interesting to test it using observations covering a wide range of column densities. Nevertheless, a more statistically significant comparison is required, especially of the simulations with the observational works reporting widespread low-velocity shocks in some clouds \citep[][]{Lefloch+98, Codella+99, JimenezSerra+10, Csengeri+11, Csengeri+16, NguyenL+13, DuarteCabral+14, Louvet+16, Cosentino+18, Cosentino+20, Liu+20_ATOMS-widespreadshocks, LiSanhueza+20_widespreadshocks, Zhu+20, Kim+23, Yang+23_CygnusX}, and here we just present the suggested test.

On the other hand, a number of observational studies have reported the presence of sub- or transonic velocity dispersions in certain regions of some filamentary structures \citep[e.g.,] [] {Arzoumanian+13, Hacar+13, Hacar+18, Friesen+16_subsonic, Suri+19, Gong+22, Koley22_subsonic, LiSanhueza+22, WangWang+23, Yoo+23_subsonic}, prompting numerical studies attempting to explain this type of motions in filaments formed by supersonic compressions \citep[e.g.,] [] {Priestley+23}. By comparison, our maps show systematically supersonic motions, consistent with the results obtained for 70~$\mu$m dark massive clumps \citep[][]{Izumi+23_ASHES}, for which Mach numbers around 4 (and up to 15) are reported. Note that the 70~$\mu$m dark cores are in extremely early evolutionary stages, and therefore feedback should have a negligible contribution in the Mach number estimate. We speculate that the subsonic motions found in certain studies may be due to the usage of tracers requiring high column densities, and therefore mainly trace compact regions along the line of sight (LOS), while the raw maps (without any excitation nor optical depth effects considered) of Fig.\ \ref{fig:sigma_N_maps} sample the full length of the 20-pc LOS. In any case, a systematic study through synthetic observations needs to be performed to settle this issue. An important consideration is that local temperature variations may affect the local sound speed, and thus introduce fluctuations in the measurements of the Mach number. For this reason, we expect the direct determinations of the nonthermal velocity dispersion to provide a more robust test.

\subsection{What is the multifractal spectrum of the dense structures?}

Since the formation of dense structures under the TS model requires strongly supersonic shocks, but not under GHC, it is likely that structures formed under the two models are fundamentally different. %In particular, structures formed by strong shocks should have a density discontinuity on one of its sides, while structures formed by gravitational contraction may have smooth boundaries as long as a singularity (a stellar object) has not formed locally.
As a consequence, the fractal dimension and the multifractal spectrum might be different for structures formed by strong shocks (i.e., within the TS scenario) and by gravitational contraction (GHC). In particular, the density multifractal spectrum is a measure of the fractal dimension of subsets of space where the mass density scales as a certain power law of size \citep[e.g.,] [] {Sreenivasan91, Chappell_Scalo01}. It is then interesting to determine the multifractal spectra in turbulence- and gravity-dominated numerical simulations \citep[e.g.,] [] {Chicana_VS01}, and compare them to those of observed clouds \citep[e.g.,] [] {Chappell_Scalo01, Elia+18, Robitaille+20}.
 
 %{\color{red} LH I don't think there is compelling evidence that clouds are truly fractal in the mathematical sense.}
 
 {%\oli EVS: Well, nothing is, because in reality the self-similar range is finite. But in practice, there is a range in which the area scales as a non-integer power of the size. And people {\it do} measure the area-size ratio as a proxy for the fractal dimension anyway. So, systematic tests of what comes out from such measurements in the two cases should be performed anyway. In regards to the multifractal spectrum, it is a more intricate diagnostic, but there are packages available for measuring it, so there's nothing to lose in measuring it.}

 %While numerous measurements of the fractal dimension of observed clouds have been performed over several decades \citep[e.g.,] [] {Falgarone+91, Turner+22, Thomasson+22, Petkova+23}, not so many
 
 %See Sun+22(Gutermuth) for fractal dimension measurements for different regions in the Serpens cloud. 

\subsection{How far does the accretion flow extend to?} \label{sec:inflow_extent}

If clouds are supported by turbulence at all but the smallest scales, as assumed in TS, then large-scale systematic inflow motions should {\it not} be present in general  {\it toward} large dense structures (clumps and filaments), since the supersonic turbulent compressions that hypothetically form them are either random and intermittent events or traveling shocks.\footnote{\citet{Padoan+20} have suggested that pure turbulent compressions can induce the observed filamentary accretion flow, but they did not show, through numerical simulations without self-gravity, that turbulence alone can produce this type of flows. We discuss this further in Sec.\ \ref{sec:inertial_inflow}.}
%which in this scenario constitute {\it post-shock} density enhancements}. At most, strong shocks should be present {\it in the surroundings} of the dense structures, as described in Sec.\ \ref{}, and a {\it low density} flow should be observed {\it upstream} of the shock (on the opposite side of the core).
Conversely, under GHC, a gravity-driven filamentary accretion flow, caused by the amplification of anisotropies expected in nearly pressureless flows \citep{Lin+65, GV14}, is expected to likely extend to the scale of the cloud itself, and to feed the cores and hubs, in a {\it conveyor belt} \citep{Longmore+14} fashion. Moreover, secondary fragmentation is expected in the filaments, forming low-mass cores with small-scale local accretion flows embedded in the larger-scale accretion flow onto the main hubs. Beyond the filament scale, GHC predicts an accretion flow {\it onto} the filaments \citep{Heitsch13, GV14}, from a (likely flattened) parent cloud \citep[see also] [] {Naranjo+22}. 

Observationally, 
%Although the 
filamentary accretion flows have been reported in many hub-filament systems
%is already well established 
\citep[e.g.,] [] {Schneider+10, Kirk+13, Peretto+14, Hacar+17, Chen+19, LiuTej+23, Seshadri+24}. A possible exception to this pattern may be the Musca filament, for which \citet{Kaminsky+23} suggested that the  observed velocity gradients are inconsistent with gravitationally-driven motions. However, their conclusion was based on comparisons with very idealized end-on filament collapse models in which important components observed in simulations where filaments form self-consistently (GV14) were missing, such as transverse accretion onto the filament from larger scales and a nearly uniform accretion flow along the full length of the filament (rather than edge-on only).

Regarding accretion onto the filaments, this has also been found in observational works \citep[e.g.,][]{Schneider+10, Kirk+13, FernandezLopez+14, Gong+18, Shimajiri+19, Zhang+20_accretionontofil, Chen+20_accretionontofil, Bonne+20_accretionontofil, Arzoumanian+21, Guo+21_accretionontofil, Gong+21_accretionontofil, Smith+23_accretionontofil, Gaudel+23_accretionontofil, Sun+24_accretionontofil}, but its spatial extent remains an open question. Even more so is the possible accretion of atomic gas onto the entire molecular clouds, with data so far being scarce and, although suggestive, not quite conclusive \citep[e.g.,] [] {Barnes+18, Heyer+22}. 
%{\cyan In this respect, it is worth mentioning a recent work by Hu, Wibking \& Krumholz (2023, arXiv:2302.11163), who find that, in galactic scale simulations, most of the molecular clouds are super-critical even when taking into account their atomic envelopes.}

%\subsection Compare the velocity gradients measured along filaments to the mass of their corresponding hubs, for a sample of hub-filament systems. There should be a relation between the mass of the hub and the velocity gradient if the motions in the filaments are produced by the gravity of the hub, as expected in the GHC scenario. 
%NOTE BY EVS: I commented out this item because the recent calculations Gilberto and I have been doing suggest that the filamentary flow far from the hubs may be similar to a rarefaction wave problem, and not directly related to the mass of the hub.

On even larger scales, \citet{Beuther+20} have presented atomic and molecular gas observations of the infrared dark cloud G28.3 and its cloud-scale environment, showing consistency with the kinematics of the gravity-driven filamentary accretion flow of (GV14), although they could not determine whether the flow around the cloud is gravitationally driven. Similarly, \citet{Zhou+24b} have reported the presence of a continuously increasing velocity gradient from large ($\lesssim 10$ pc) to small ($\sim 0.1$ pc) scales in the G333 giant molecular cloud complex, consistent with gravitational acceleration throughout this scale range.

\subsection{Are the typical mass of fragments and their typical separations given by the thermal Jeans mass and length?
%Is the typical mass of fragments and their typical separation changing as parental cores evolve?
%Are super-Jeans cores systematically younger than nearly-Jeans ones?
} 

A common strategy for studying the fragmentation of massive dense cores is to compare the masses of their fragments to the Jeans mass of the cores. 
While in some cases the typical masses of most of the fragments have been reported to be of the order of the Jeans mass of their parent structure, $\MJ$ \citep[e.g.,][]{Palau+15, Cyganowski+17, Henshaw+17_lowmassfragments, Heyer+18_fragmentation, Beuther+19_Ljeans-sub50au, Sanhueza+19, LuCheng+20_fragmentation, Palau+21, Walker+21_fragmentation,
Saha+22_fragmentation, 
Morii+24_ASHES}, in other cases the fragment masses are observed to be significantly larger, and these are referred to as {\it super-Jeans} fragments \citep[e.g.,][]{Wang+14_superJeans, Zhang+15_superJeans, Figueira+18, Liu+20_superJeans, Barnes+21}.\footnote{The works of \citet{Zhang+09_turbsupport}, \citet{Pillai+11}, \citet{Wang+11_turbsupport}, \citet{Lu+15_turbsupport}, \citet{Feng+16_turbsupport}, \citet{Csengeri+17}, and \citet{Sanhueza+17_turbsupport}, also reporting super-Jeans fragments, do not reach enough mass sensitivity to detect masses of the order of the Jeans mass, and could be missing an important population of low-mass fragments consistent with thermal Jeans values.} A frequent interpretation of the super-Jeans fragments is that the parental cores are fragmenting into substructures (the ``fragments'') of the order of the ``turbulent Jeans mass'' $\MJt$. The latter is most frequently estimated by replacing the sound speed in the standard (thermal) Jeans mass by the total velocity dispersion \citep[e.g.,][]{Zhang+09_turbsupport, Pillai+11, Wang+11_turbsupport, Wang+14_superJeans, Lu+15_turbsupport, Traficante+23}. Note, however, that, even within the context of TS, the most accurate interpretation of the turbulent Jeans mass must consider that, besides having a larger-than-thermal velocity dispersion, the supersonic turbulence induces density fluctuations, of amplitude $\sim \Ms^2$, where $\Ms$ is the sonic Mach number, if the gas is nearly isothermal. This density enhancement must be accounted for in order to determine (in)stability. Therefore, the turbulent Jeans mass\footnote{It is important to note that the `turbulent Jeans fragmentation', as considered in equations \eqref{eq:turb_MJ} and \eqref{eq:turb_MJ_pract}, and also referred to in some works as `gravo-turbulent fragmentation', should not be confused with plain `turbulent fragmentation', in which the density fluctuations are produced exclusively by turbulence and gravity plays no role \citep[e.g.,][]{VS94, Pad_Nord99, Passot_VS98}.} should be taken as \citep{MacL_Kles04,Palau+15}
\begin{equation}
\MJt = \frac{\pi^{5/2}} {6} \frac{\left[(\Ms^2+1) \cs^2\right]^{3/2}} {G^{3/2} (\Ms^2 \rho)^{1/2}},
\label{eq:turb_MJ}
\end{equation}
where $\rho$ is the mean density of the core. In practical units, and neglecting the thermal support,
\begin{equation}
\Big[\frac{M_\mathrm{J,t}}{M_{\sun}}\Big] = 
0.8255\,\Big[\frac{\sigma_\mathrm{1D,nth}}{0.188\,\mathrm{km\,s}^{-1}}\Big]^3
\Big[\frac{n_\mathrm{H_2}\,\Ms^2}{10^5\,\mathrm{cm}^{-3}}\Big]^{-1/2},
\label{eq:turb_MJ_pract}
\end{equation}
where $\sigma_\mathrm{1D,nth}$ is the non-thermal velocity dispersion along the line-of-sight, and $n_\mathrm{H_2}$ is the density of H$_2$ molecules \citep{Palau+15}.

On the other hand, under GHC, several points must be considered in the interpretation of the super-Jeans fragments. First, it is is important to note that the Jeans mass used to compare the fragments' masses should be the one corresponding to the time when the fragments first decoupled from the core and began to contract on their own, when the core's density was lower, and thus its Jeans mass was larger, as proposed by \citet{VS+19} and recognized, for example, by \citet{Csengeri+17} and \citet{Xu+24_ASSEMBLE}.

Second, the existence of strongly super-Jeans fragments can be interpreted simply as a consequence of the growth of the fragments' mass by accretion. Therefore, they should eventually undergo sub-fragmentation themselves, except if the fragment's Jeans mass has begun to increase by local heating, either by stellar heating or by an increase of the opacity.

Third, 
%whose fragment masses are significantly larger than the core's Jeans mass may be interpreted as indicative of the cores being in an early evolutionary stage, because 
the evolution and growth of the fragments implies a delay between the time when a fragment begins to contract locally, and the time when it appears as a clearly identifiable substructure---i.e., with a sufficient density contrast over its parent structure. This is a consequence of the facts that {\it a)} the fragment must accumulate at least a few Jeans masses in order to begin fragmenting itself; {\it b)} even though a sub-fragment may have already begun to collapse on its own, the initial stages of the collapse proceed very slowly \citep[see, e.g.,] [] {Girichidis+14, VS+19}, and thus it may not be clearly distinguishable from its parent structure (the fragment) until a substantial fraction of its free-fall time has passed. Therefore, even if the fragment has begun to sub-fragment, this may not be immediately observable; {\it c)} finally, the sub-fragmentation scale may be below the resolution of the currently available observations, especially for very distant regions. This kind of evolutionary effects in the fragmentation process has already been proposed by \citet{Louvet+19_littlefrag} to explain the small amount of fragmentation found in a sample of starless massive dense cores. These authors suggest that the cores with no compact objects have probably initiated the fragmentation process, but the fragments did not have time to accrete and be detectable yet.

In summary, under GHC, it is expected that a significant fraction of strongly super-Jeans fragments may exhibit internal sub-fragmentation, but the observation of this further level is challenging. For this reason, the comparison of the fragment masses with the parent cores' Jeans masses is not a straightforward test.

On the other hand, 
%{\cyan A possible test would be to study the relation between the ratio of $M_\mathrm{frag}$/$M_\mathrm{Jeans}$\footnote{$M_\mathrm{frag}$ is the average mass of the fragments embedded in a massive dense core, and $M_\mathrm{Jeans}$ is the Jeans mass of the parental core given by its average density and temperature.} vs the luminosity-mass ratio of the parental core, a well-known evolutionary indicator \citep{Molinari+08}. However, this trend could be reverted by the fact that fragments are continuously accreting and gaining mass. For example, \citet{Xu+24_ASSEMBLE} find evidence that the mass and surface density of the fragments actually increases with time due to continuous accretion. While the evolution of the average masses of the fragments is not immediately related to the parental core evolution, the separations are obviously not affected by continuous accretion and should more clearly show a decrease with time.
%{\mst{Similarly,} Further evidence of the ongoing gravitational contraction of the cores is provided by} 
a cleaner test may be provided by the mean separation between the fragments, which is in principle given by the Jeans length in the parent core when the fragmentation occurred, and is not complicated by the fragment growth by accretion.
Many observational works have reported that the measured separations are of the order, or smaller, than the Jeans length \citep[e.g.,][]{Takahashi+13, Teixeira+16, Kainulainen+17_fragmentation, Ohashi+18, Beuther+18, Palau+18, Beuther+19_Ljeans-sub50au, LiMyers+19, LiuChen+19_fragmentation, Svoboda+19, Sanhueza+19, LuCheng+20_fragmentation, Zhang+21, Walker+21_fragmentation, Saha+22_fragmentation, Morii+24_ASHES, Ishihara+24}.
The expectation under GHC in this case is that the typical separation between fragments should in general be of the order or less than the Jeans length of the parent core, because, by the time the fragments are distinguishable, they may have approached one another due to global core contraction, as suggested by, e.g., \citet{Takahashi+13}, \citet{Teixeira+16} and \citet{Xu+24_ASSEMBLE}. A corollary of this expectation is that samples of more evolved massive dense cores should exhibit smaller typical fragment separations than younger samples, as has been recently reported by, e.g., \citet{Traficante+23} and \citet{Xu+24_ASSEMBLE}. 

%report that the typical separations between fragments decreases with time, consistent with the Jeans length decreasing in time due to the increase in density of the parent structures. In addition, \citet{Xu+24_ASSEMBLE} also estimate that the observed separations between fragments are consistent with an earlier Jeans length and the subsequent contraction of the parental core.

%Their evolutionary stage may be probed by the age of the protostellar objects they contain, and so a test of whether these objects are evolving by gravitational rather than turbulent fragmentation is that the most super-Jeans cores should {\mst{be younger} contain younger protostellar objects, or none at all}. 

%{\mgt In addition, the fact that the Jeans mass decreases with time in the cores implies that the typical mass of the fragments decreases with time.} {\cyan This is consistent with the recent work reported by \citet{LiLiu+23}, where unambiguous evidence was found that more low-mass stars are progressively formed with time in the Solar neighborhood, that is, that the IMF is top-heavy at early times. 

\subsection{Are more actively star-forming regions (in a given mass range) older?}

The GHC model predicts an increase of a molecular cloud's star formation activity as the cloud evolves by accretion of external material and global gravitational contraction \citep[e.g.,] [] {ZA_VS14, Camacho+23}. However, as discussed in the latter reference, the evolution is parameterized by the total initial mass participating in the collapse, so that, as a first approximation, the evolution of accreting molecular clouds can be determined by two parameters, age and instantaneous mass, or proxies for each on of them. This was used by \citet{VS+18} to interpret the scatter in the observed instantaneous SFE (cf.\ eq.\ [\ref{eq:SFE}]) of the cloud sample of \citet{Lada+10} as a consequence of an age spread among those clouds.

Therefore, a test of the evolution predicted by GHC is whether clouds with larger SFRs in a given mass range should contain older YSOs than those with lower SFRs. It is important to emphasize, though, that, in order to perform this test, the clouds should be binned per mass intervals, as more massive clouds in earlier evolutionary stages could have a similar SFR as a more evolved but less massive cloud.

\subsection{The brown dwarf quest: spatial distribution and temporal appearance of brown dwarfs in molecular clouds} \label{sec:BD_quest}

The formation of brown dwarfs (BDs) remains a matter of intense debate \citep[e.g.,] [] {Luhman12, Chabrier+14, Offner+14, Lomax+16, Whitworth18, Palau+24_review}. In particular, an important unanswered question related to this topic is whether the core mass function (CMF) is causally related to the stellar initial mass function (IMF). This problem arises because, as the gas collapses to form a star, the local density increases, and thus the Jeans length and mass continuously decrease for as long as the gas remains isothermal. More importantly, for isothermal gas, the Jeans length of a fixed-mass collapsing gas parcel decreases faster than the parcel's size \citep{Hoyle53}, and therefore the parcel can continue to fragment into ever-smaller pieces, leading to the process of {\it gravitational (or Hoyle) fragmentation}. This author also proposed that the fragmentation process is only expected to stop when the gas parcel becomes so dense that it becomes optically thick, and begins to trap the heat generated by the collapse. At this point, the effective thermodynamic behavior reverts from nearly isothermal to nearly-adiabatic, eventually halting the collapse \citep{Hoyle53} and forming a {\it first hydrostatic (or Larson) core} \citep{Larson69}, of mass $\sim 0.003$--$0.05 \Msun$ \citep{Boyd_Whit05, Young23_FHCs}. From then on, the mass of this ``seed'' Larson core grows by accretion. At this point, the accretion may continue until the whole mass of the molecular core is fully accreted, as assumed in the models that relate the stellar masses to their parent core masses \citep[e.g.,] [] {Pad_Nord02, Henneb_Chab08}, or it may be regulated by the mechanism of {\it competitive accretion} \citep{Bonnell+01}, in which the various seeds chaotically compete for the available material in the surroundings.

Under TS, the implicit assumption is frequently made that the accretion exhausts the available gas reservoir, or that it is consumed up to a certain typical efficiency. For example, theories of the IMF within this context usually attempt to derive the CMF as a proxy for the IMF \citep[e.g.,] [] {Pad_Nord02, Henneb_Chab08, Chabrier_Henneb11, Hopkins12}, and the {\it turbulent core} model \citep{McKee_Tan03} assumes that a massive core must form and collapse monolithically to form a massive star. In this context, \citet{Pad_Nord04} have suggested that the formation of BDs requires particularly strong shocks, of Mach number $\Ms \sim 10$, which produce high enough densities for the Jeans mass to reach substellar values.\footnote{Note, however, that \citet{Lomax+16} have pointed out that rather than very high Mach numbers, the requirement appears to be a highly focused turbulent compression, and that this appears unlikely. Therefore, they concluded that this may not be the dominant BD formation mechanism.}

Instead, the GHC scenario is fundamentally based on the Hoyle gravitational fragmentation scenario, in which a molecular cloud core is likely to continue fragmenting down to masses of the order of the first Larson core. Therefore, a significant number of seeds must eventually form within a molecular cloud core, being all embedded in a core-scale accretion flow. These seeds must then compete for the accreting material, locally developing a competitive accretion process, with the subtlety that continuous replenishment of material should be taken into account under GHC, and considering also that dynamical interactions might happen among the different seeds. That is, as stated in \citet{VS+19}, GHC can be viewed as a large- (cloud-)scale scenario in which the small- (core-) scale evolution is described by the competitive accretion model with continuous mass replenishment and dynamical interactions, as shown in some simulations of \citep[e.g.,][]{Reipurth2015_BinaryBD, Bate2019}. In the latter, BDs are thought to form as the the most disfavoured objects in the competition for the accreting material. Therefore, no strong shocks are necessary under GHC to form BDs, and thus such shocks are not necessarily expected near nascent BDs in this scenario.

\subsection{Does the virial parameter of cores and clumps evolve with time?}

In the TS scenario MCs are supported by turbulence against gravity, and therefore they must be in a quasistatic state. Therefore, no systematic evolution is expected for them according to TS. Instead, GHC is an intrinsically evolutionary scenario, because the clouds' masses and mean densities increase with time due to accretion, if the clouds are defined in terms of density, column density or intensity thresholds. This increase has been consistently observed in numerical cloud and clump samples in both SPH \citep{Camacho+20, Camacho+23} and AMR simulations \citep{Colin+13}. Specifically, the location of the clouds, clumps and cores traces specific trajectories in the \LS\ and \aM\ diagrams as they evolve, while simultaneously their SFR increases. Moreover, \citet{Camacho+20} compared the evolution of the star-forming regions in the simulations to clouds in different evolutionary stages, showing good agreement in all parameters. However, there are few observational evolutionary studies of the virial parameter or the Larson ratio \citep[e.g.,] [] {DeBuizer+23}, and a statistically-significant number of such studies is still lacking.

%{\cyan Camacho+20,+23 and De Buizer+23 find that the virial parameter evolves with time, while in TS this is not expected.}

\section{Discussion} \label{sec:disc}

\subsection{Underlying theoretical models and interpretation of simulations} \label{sec:underl_theo_mods}

An important point to notice is that numerical simulations large enough to include the necessary spatial and temporal extensions (at least tens of parsecs over tens of megayears) to allow the formation and evolution of clouds, but not so large that individual clouds are unresolved or averaged over, all share similar physical ingredients; in particular, self-gravity, radiative cooling, and various forms of stellar feedback. This is the case of simulations such as those by \citet{Hill+12}, \citet{Colin+13}, \citet{IbanezM+16}, \citet{Padoan+16},\footnote{Note, hoever, that the simulations by \citet{Padoan+16} include supernovae in a closed box of 250 pc, and thus do not allow the large-scale vertical flows that may relax the high pressure generated by the SNe, and thus may be overdriven.} \citet{ZA+19}, \citet{Colman+22}, and the {\sc Silcc} \citep[e.g.,] [] {Walch+14}, {\sc Tigress} \citep[e.g.,] [] {Kim_Ostr17, Kim+20} and {\sc Starforge} \citep[e.g.,] [] {Grudic+21} collaborations. However, the interpretation of these simulations can be radically different, depending on the underlying theoretical model assumed to describe the processes, which in turn determines the kinds of tests performed on the simulations. A similar situation applies for observational studies. 

The most notable example of this situation is that, when the TS scenario is assumed to be the appropriate description of the clouds, the linewidth-derived velocity dispersion is interpreted as corresponding exclusively to turbulence, which opposes gravity. In this case, for example, a value of the virial parameter $\av \sim 2$ is generally interpreted as ``marginal gravitational binding'', as it is thought to correspond to weaker binding than virial equilibrium, for which $\Ek \sim |\Eg|/2$ \citep[e.g.,] [] {McKee_Ostr07}. Instead, under GHC, $\av \sim 2$ is regarded as representative of gravitational collapse, since it corresponds to $\Ek \sim |\Eg|$ \citep{BP+11, BP+18}. Therefore, tests need to go beyond the simple measurement of the kinetic and gravitational energies, and search to determine whether the motions correspond to organized infall or to random turbulence, and in what proportions \citep[e.g.,] [see Sec.\ \ref{sec:linewidths_turb_coll} for further discussion] {GG_VS20}.

Similarly, suggestions that clouds or their substructures are unbound have been made on the basis of them having values of $\av$ significantly larger than 2. Although this is true in a strict sense {\it locally}, it is nevertheless possible that these locally unbound objects may be just the ``tips of the iceberg'' of larger-scale structures that themselves {\it are} bound and collapsing \citep[e.g.,] [] {Gomez+21, Camacho+23}, and so the small, unbound objects may be undergoing compression from the infall of the surrounding material. Thus, the boundedness of objects needs to be tested in the context of their environment, rather than in isolation.

\subsection{Comparison with other models} \label{sec:other_models}

In this section we discuss several recently proposed models that fit to a certain degree within TS or GHC, in spite of having been proposed as distinct models.

\subsubsection{The inertial inflow scenario} \label{sec:inertial_inflow}

Within the context of the TS scenario, \citet[] [herafter P20] {Padoan+20} have proposed that pure supersonic turbulence is capable of producing hub-filament systems (HFSs) similar to those detected observationally, and have referred to this property as the {\it Inertial Inflow} (I2) model, as it assumes that the turbulent compressive motions forming the HFSs are of pure inertial nature. P20 based their suggestion on the statement that, in clouds at the (large) scale of a few to several parsecs, ``the kinetic energy of the turbulence [...] usually  exceeds both the thermal energy (velocities are supersonic) and the gravitational energy''. 
%than gravitational energies \citep[e.g.,] [] {Leroy+15, MivilleD+17, Traficante+18a}, and are thus dominated by turbulence rather than gravity. 

However, this line of argument can be questioned on several flanks. First, clouds that tend to be unbound are not necessarily the largest, as assumed by \citet{Padoan+20}. Rather, they are the low-column density objects in the KH diagram, or the low-mass ones in the \aM\ diagram. For example, \citet{Camacho+23} have pointed out that, in the KH diagram, there is an ambiguity at low column densities, since a low-column density object can be either weakly bound if it is small (and therefore, of low mass, since $\Sigma = M/\pi R^2$; that is, the ``chaff''; cf. Sec.\ \ref{sec:devs_vir}), or strongly bound if it is large (and therefore, massive). This ambiguiyty is clearly illustrated by the wide range of the Larson ratio in low-column density objects seen in many studies presenting the KH diagram \citep[e.g.,] [] {Leroy+15, MivilleD+17}. The same happens in the \aM\ diagram at low masses.\footnote{For this reason, \citet{Camacho+23} have suggested that a convenient procedure to avoid this ambiguity is to plot directly the kinetic energy versus the gravitational one.}
%In particular, in the latter diagram, the trend is repeated for different classes of objects \citep[cf.\ Figure 1 of] [] {Kauffmann+13}, from giant molecular clouds to early-stage massive clumps to dense low-mass cores, and thus the appearance of this trend is not a matter of size scale, but rather low gravitational binding at any scale. 
Note that this also implies that, at any given column density, it is the most massive, and therefore {\it the largest}, objects that are more strongly bound, as can be inferred from Figure 1 of \citet{Kauffmann+13} and several other cloud surveys \citep[see, e.g., Fig.\ 2 of] [and references therein] {Chevance+23}, and verified in numerical simulations by \citet{Camacho+16}. 

Second, although most of the low-column density clouds in the study by \citet{Evans+21} have large virial parameters, these are the {\it least massive clouds in their sample}. However, by mass, at least half of the mass in their sample may be gravitationally bound, since most of the mass is in the less numerous, most massive clouds \citep{Williams_McKee97}. 

Finally, it must be pointed out that P20 based their argument of kinetic energy domination at the (relatively large) scale of HFSs on the results of simulations that {\it did} include self-gravity \citep{Padoan+16}, and only showed a schematic diagram {\it suggestive} of the mechanism (see right panel of Fig.\ \ref{fig:HFS_num}), rather than showing that simulations {\it without self-gravity} can produce such systems. Without this test, it cannot be asserted that, in reality, those systems in their simulations are not actually generated by self-gravity, and that an important fraction of the kinetic energy in those systems may have a gravitational origin. In fact, producing the highly focused flow patterns towards the filaments and the hubs, as observed in HFSs, does not appear feasible with pure turbulence, given the random nature of the turbulent compressions. P20's suggestion was only an {\it inference}, arising from the observed supervirial nature of many low-column density structures in the HK diagram, which they interpreted as being a property of large-scale structures, while in reality it is a property of {\it small}-scale structures in that column density range. The large-scale (and massive) structures in that column density range are the most strongly gravitationally bound \citep{Camacho+16}.

%P20 suggest that the filaments are the post-shock regions generated by inertial, large-scale supersonic turbulent compressions (see Fig.\ \ref{fig:HFS_num}). Thus, consistent with the TS framework, the I2 model assumes that star-forming clouds are not undergoing cloud-scale gravitational contraction, but are rather just turbulent fluctuations themselves in a globally turbulent medium \sout{{\cyan that effectively offers support against gravity}} {\mgt and are themselves supported (or dispersed) by their internal turbulence}. {\mst{The I2 model then suggests that hub-filament systems can be directly created by the inertial compressions.}}

\begin{figure*}
\centering
\vcenteredhbox{\includegraphics[width=0.45\linewidth]{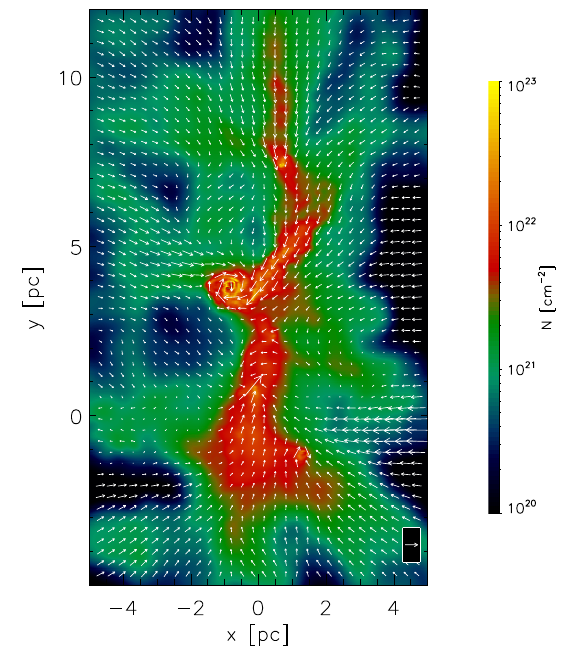}}
\vcenteredhbox{\includegraphics[width=0.45\linewidth]{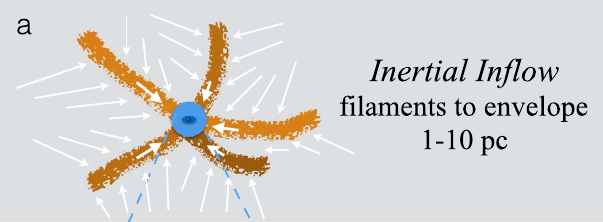}}
\caption{{\it Left:} Column density (color scale) and velocity field (arrows) projected onto the $(x,y)$ plane from a hub-filament system arising in a numerical simulation of a cloud undergoing global, anisotropic and hierarchical gravitational contraction (GV14). Shown are the central hub (with a rotating motion) and the filaments converging to it, which have secondary collapsing sites (cores). The arrow at the lower-right inset indicates velocity of 2$\kms$. {\it Right:} Schematic diagram of the assumed hub-filament inertial flow that would be produced by inertial turbulent compressions \citep[from] [] {Padoan+20}. The accretion flow depicted by this diagram arises spontaneously in simulations of clouds undergoing GHC, while no pure-turbulence simulations, without self-gravity, have been shown to produce it.}
\label{fig:HFS_num}
\end{figure*}

%showing that the transverse accretion flow from the cloud to the filament, and the longitudinal accretion flow from the filament to the hubs can develop purely by turbulent compressions. In P20, only a {\it schematic} diagram suggesting so is shown (right panel, Fig.\ \ref{fig:HFS_num}). {\mst{Since the simulation presented by P20 is dominated by supernova-induced turbulence, we speculate that, if hub-filament systems form at all, they will only occur at scales quite smaller than those of real ones, where gravity does dominate. It also}. It} remains to be shown that hub-filament systems {\mgt with the observed kinematic structure} arise at all in simulations without self gravity.

On the other hand, under GHC, HFSs are formed by cloud-scale anisotropic gravitational contraction (see left panel of Fig.\ \ref{fig:HFS_num}). However, during this process, 
%In fact, no proof was provided by P20 that the {\mst{hypothetical inertial}} accretion flow onto the filaments in their simulation is not due to the gravitational potential of the filament's {\it parent} cloud{\mgt , which is the proposed mechanism in GHC}. Indeed, within the context of GHC, 
it is expected that {\it any} density structure in its early stages of development will appear to be {\it locally} dominated by the kinetic energy, because its mass has first to grow sufficiently by accretion before its own self-gravity can dominate \citep{BP+18, VS+19}. Nevertheless, the converging flows that are compressing and assembling it may be the result of the gravitational contraction of a larger-scale structure, which then compresses its innermost regions, since the collapse actually occurs from the outside-in \citep[e.g.,] [] {Gomez+07, Gomez+21, Gong_Ostr09, Ganguly+22, Camacho+23}. That is, the first stages of compression of any structure cannot be due to the structure's own self-gravity---which is necessarily very weak initially---, but must be triggered from the outside. However, if this external triggering is due to the infall of larger-scale collapsing structure onto its central parts, then the latter can be considered simply as the not-yet-self-gravitating ``tip of the iceberg'' of the former. Of course, a purely inertial origin of the compression, such as that due to the collision of expanding bubbles, is also possible. Thus, an important point to elucidate is the fraction of the total molecular gas mass that is gravitationally bound {\it at the largest scales}.
%{\cyan AINA: IT IS NOT CLEAR TO ME THAT THIS HAS TO BE NECESSARILY TREATED AS A SEPARATE MODEL. COULD WE CONSIDER TO INCLUDE THE I2 MODEL ALREADY AT THE BEGINNING, OR EVEN CALLING THE TS-I2 SCENARIO?} \gil{It is mentioned that it should be considered as part of TS. Also, specific objections to this scenario are presented here. I second to keep the separate discussion.}

\subsubsection{The turbulent core model} \label{sec:turb_core}

\citet[] [the latter hereafter referred to as MT03] {McKee_Tan02, McKee_Tan03} presented a model that has become a reference for the description of the rapid formation of massive stars within turbulent, massive dense cores. This model is commonly referred to as the {\it turbulent core} (TC) model, and predicts the formation of massive stars on timescales $\sim 10^5$ yr, necessary for consistency with observationally inferred timescales. 

The model accomplishes this by considering massive ($M \gtrsim 10^3 \Msun$), high-column density ($\Sigma \sim 5 \times 10^3 \Msun$~pc$^{-2}$, or $N \sim 2.5 \times 10^{23} \psc$) cores of sizes $\sim 0.5$ pc, in equilibrium with their turbulent, magnetic and/or thermal pressure. The large pressure implies a large effective ``sound'' speed $\ceff$, which in turn implies a high protostellar accretion rate, given by
\begin{equation}
\dot M_* \approx \frac{\ceff^3} {G},
\label{eq:accr_rate}
\end{equation}
thereby allowing for the required short timescales.

It is worth noting that this model assumes that the cores are supported by supersonic turbulence and/or magnetic fields, and then considers the effective signal transmission speed implied by the total pressure. Therefore, it formally makes the same assumption of global support that the TS model makes for entire clouds, but applied at the core scale, and implies that the core has a lifetime much longer than its own free-fall time. This assumption is essential to the model in order to explain the existence of YSOs of ages several times longer than the local free-fall time, as well as the low efficiency of star formation. 

In order to conform to the long lifetime of the core, MT03 assume that most of the turbulent density fluctuations inside the core are Jeans {\it stable}, and do not proceed to collapse individually, implying that the core does not fragment but rather proceeds to collapse as a whole, forming a single massive star or perhaps a massive binary or multiple star (see their Sec.\ 3). This implication has prompted intense searches for high-mass starless cores \citep[i.e., cores with no star formation activity such as infrared sources or molecular outflows; e.g.,][]{Motte+18_review, Barnes+23_HMSC, Morii+23_ASHES}, which in most cases, however, have not been successful.

Alternatively, under GHC the cloud, clump, and core masses systematically increase in time by accretion, until they are disrupted by stellar feedback \citep{VS+10, VS+17, VS+24, Colin+13, MacLow+17, GS_VS20, Gomez+21}. This implies that clumps and cores begin forming stars a few megayears before they themselves become massive enough to form massive stars \citep{VS+17, VS+24}, and so a population of low-mass YSOs is already expected by the time massive stars begin forming in the core (by then, a hub). So, isolated high-mass startless cores are not generally expected under GHC.

It is worth noting that the main goal of the TC model was to provide a large protostellar accretion rate that would allow the formation of massive stars in a timescale $\sim 10^5$ yr, for which a large effective sound speed is necessary. They accomplished this by assuming a large turbulent velocity dispersion (which must be confined to scales much smaller than the core size in order to provide an isotropic pressure). However, the same effect can be obtained from gravitational collapse, since the infall speed is very similar to the virial speed (cf.\ Sec.\ \ref{sec:Larson_rels}), and therefore the ram and turbulent pressures are of similar magnitude. The advantage of the GHC perspective is that replacing the assumption of virial equilibrium by that of collapse flow and core growth by accretion removes the problematic implication of the TC model that high-mass starless cores should be as common as high-mass stars, which observations do not support.

%However, aside from the implication of prestellar massive cores, the dynamical features required for the TC are similar to those predicted by GHC, because the free-fall speed is very similar to the virial speed (cf.\ Sec.\ \ref{sec:Larson_rels}), and provides an infall ram pressure comparable to the turbulent pressure assumed in the TC model.} Therefore, the TC model could be fitted into the GHC perspective if only the assumption of turbulent virialization were replaced by the assumption of gravitationally-driven inflow. This is consistent with the facts that {\it a)} \citet{McKee_Tan03} acknowledge that the timescale for massive-star formation is of the order of the free-fall time of the {\it parent cloud} embedding the massive core, similarly to the prediction of GHC \citep{VS+19}, and {\it b)} accretion is now routinely observed to occur at least up to the clump (parsec) scale along filaments (cf.\ Sec.\ \ref{sec:inflow_extent}). 

%{\cyan However, a prediction from this model is the existence of high-mass starless cores and they have been very elusive to find so far (see, e.g., the review by Motte+18 and other REFS). FROM AN OBSERVATIONAL POINT OF VIEW, IT IS VERY DIFFICULT TO THINK THAT THE GHC AND TC ARE EQUIVALENT.}

\subsubsection{The competitive accretion scenario} \label{sec:comp_accr}

Since the turn of the century, the main alternative to the TC model for massive-star formation (and, in fact, the development of the entire stellar mass range) is the {\it competitive accretion} (CA) scenario \citep{Bonnell+01, Bate+03}. As pointed out in Sec.\ \ref{sec:BD_quest}, the GHC scenario is essentially similar to the CA scenario, except that it assumes that the same phenomenology extends to the scale of entire MCs, and therefore focuses on the evolution of the clouds, their SFR, and the formation of their substructures, such as hub-filament systems, rather than on the stellar mass distribution. But the physical processes involved (multi-scale accretion and gravitational fragmentation) are essentially the same.

\subsubsection{The hub-filament system scenario} \label{sec:hub_fil}

From the analysis of observations from the {\it Herschel} survey {\sc HiGal}, \citet[] [hereafter K20] {Kumar+20} have recently proposed a unified {\it hub-filament system paradigm} (HFSP) for star formation based on the development of hub-filament systems in molecular clouds. In this scenario, the filaments feed the hubs, in such a way that low- and intermediate-mass stars form in the filaments, while high-mass stars can form only at the hubs, where several filamentary structures intersect, due to the enhanced density they acquire by accretion from the filaments. K20 suggest that the spatial and temporal separation of the regions forming low- and high-mass stars generates the mass segregation and age spreads observed in young clusters. 

Although K20 compare their model to GHC, as well as to the ``conveyor belt'' model advanced by \citet{Longmore+14}, and point out some differences with both of them, most of the differences are unsubstantial, such as the systems that are used as analogies in VS19 and K20 (rivers driven by a gravitational potential difference in GHC and electric currents driven by an electric potential difference in the HFSP), and that while in GHC all massive clumps form massive stars, only those in HFSs do so in the HFSP. The latter ``difference'' is really nonexistent, as in GHC massive clumps are systematically hubs that are fed by filaments (GV14). 

More importantly, K20 argue that the HFSP simply assumes that clouds are filamentary in structure and that there is no collapse at the cloud scale, as in GHC. Unfortunately, K20 do not discuss the plausible origin of the HFSs and, if they rule out global cloud-scale gravitational contraction by assumption, the only alternative mechanism is supersonic turbulence. In this sense, the HFSP might be considered equivalent to the I2 scenario. However, as discussed in Sec.\ \ref{sec:inertial_inflow}, it has not been shown through numerical simulations that turbulence alone (i.e., without gas self-gravity) can generate the organized filament-to-hub flow that the HFSP itself requires.

In conclusion, rather than considering GHC and the HFSP as competing scenarios, we view them as essentially the same scenario, the HFSP arrived at from the observational viewpoint and GHC arrived at from the numerical simulation viewpoint.

\subsubsection{The cloud-cloud collision scenario} \label{sec:CCCs}

In recent years the cloud-cloud collision (CCC) scenario, initially proposed by \citet{Furukawa+09}, and recently reviewed by \citet{Fukui+21}, has received considerable attention. This scenario proposes that massive-star forming regions are the result of inertial collisions between clouds that strongly compress the gas, promoting copious star formation, and allowing for massive-star and cluster formation. This scenario has also been invoked by a number of works as the formation mechanism of hubs or intersecting filaments \citep[e.g.,] [] {Beltran+22_cloudcloudcollisions}, and simulated by numerous groups \citep[see] [and references therein] {Fukui+21}.

%there has been a significant increase in the number of works invoking the cloud-cloud collision scenario to explain observations. From the theoretical side, recent works such as \citet{Hsu+23} have studied the Core Mass Function expected in this scenario.

The observational signatures that have been interpreted as indicative of collisions between unequal-size clouds are \citep{Fukui+21}: {\it a)} A ``complementary'' structure in an intensity map, such that two regions with clearly distinct velocities appear to be two parts of the same observed intensity structure. {\it b)} A ``bridge'' feature in a position-velocity (PV) diagram that connects the two different-velocity clouds. {\it c)} Sometimes, a ``U'' shape of the combined structure in the intensity map. It is important to note, however, that often the two colliding ``clouds'' are described as ''filaments'' \citep[e.g.,] [] {Fukui+15} or ``hub-filament systems'' \citep[e.g.,] [] {Maity+22}. 

It is also important to note that the observational signatures of the proposed cloud-cloud collisions are not qualitatively different from those expected from the infall of material from filaments onto a hub, as proposed in the GHC scenario and the HFS scenario of \citet{Kumar+20}. Indeed, HFSs can be viewed as a ``conveyor-belt'' type of flow \citep{Longmore+14}, in which material is funneled from the large to the small scales via an accretion flow that proceeds from cloud to filament and from filament to hub (GV14). Indeed, Fig.\ \ref{fig:fil_projs} shows 4 projections in the position-velocity space of one of the filaments studied by GV14, where the two components with different velocities are seen to be connected by a bridge-like feature, which is more prominent at larger projection angles.

\begin{figure*}
\centering
\includegraphics[width=\linewidth]{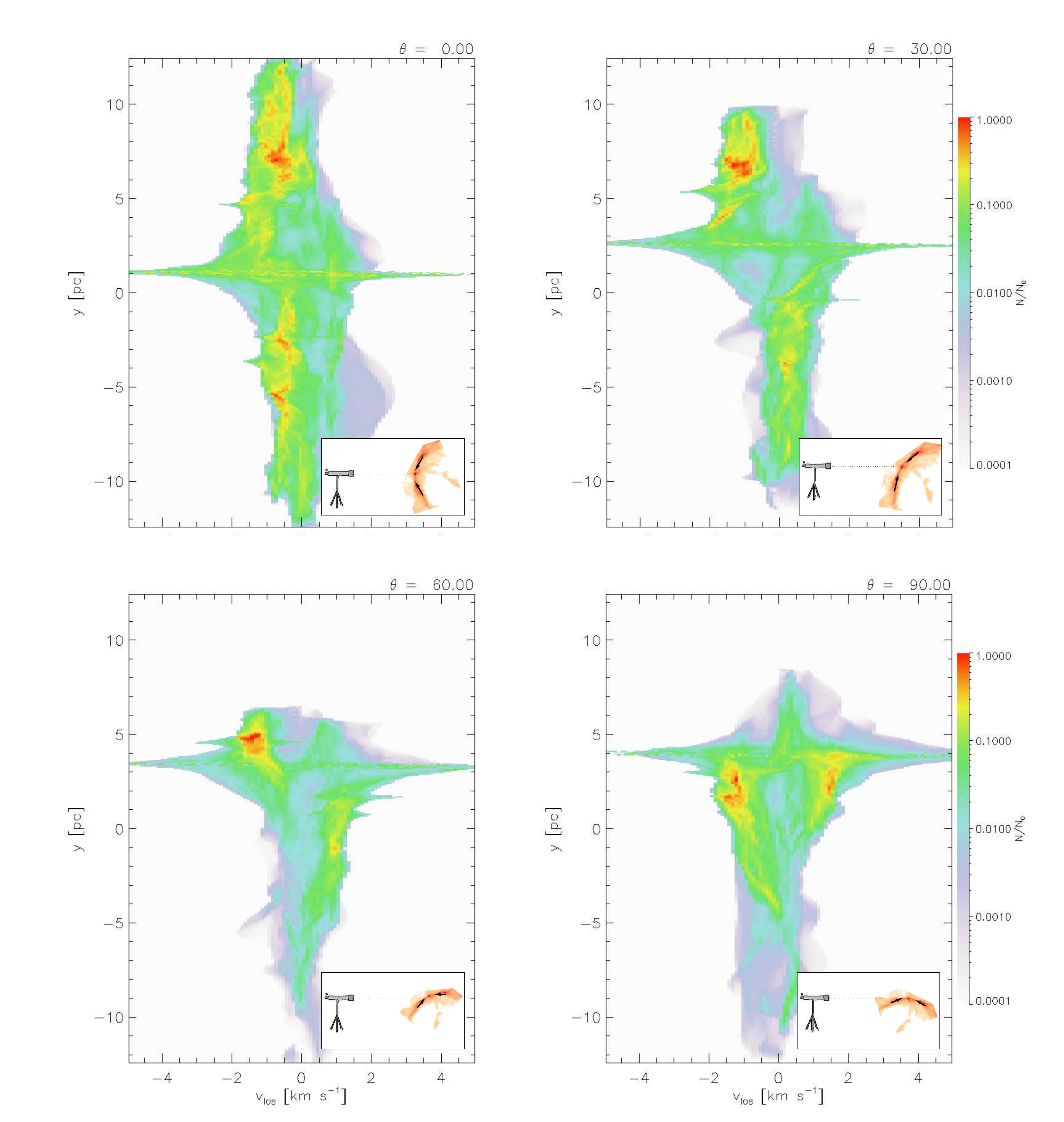}
\caption{Four projections in PV space of one of the filaments arising through anisotropic gravitational contraction, leading to cloud-to-filament and filament-to-hub accretion flow, in the GHC simulation studied by GV14. The $y$-coordinate indicates the direction along the axis of the projected, inclined filament. The projection angle with respect to the tangent to the filament at the hub are indicated at the upper right corner of each panel. Note the velocity difference between the two (upper and lower) branches of the filament and the large velocity dispersion at the hub near the middle of the images. Note also that the two branches seen to extend downwards from the hub at large projection angles because the filament is actually curved (arc-like), curving backwards from the plane of the top left panel, and downwards from the hub at the bottom right panel. These features are similar to those attributed to cloud-cloud collisions, since the filament has a U-shape, and the PV-plot shows a V-shape \citep{Fukui+21}.}
\label{fig:fil_projs}
\end{figure*}

Therefore, under GHC, the conveyor-belt process generates a single dense hub in which two or more components of distinct velocities meet, and appear clearly separate in a PV diagram. Moreover, depending on the inclination, the filamentary extensions may be more or less noticeable. We therefore suggest that the cloud-cloud collision scenario can be interpreted as a manifestation of the HFSs that arise due to gravitationally-driven flow during GHC.

Finally, we also note that the possibility of dense clouds moving ballistically through the intervening lower-density medium is not consistent with the fact that clouds are simply the ``tips of the iceberg'' of a continuous density distribution, and are not expected to move very quickly with respect to the diffuse gas. Instead, they are formed at the {\it stagnation points} of the flow \citep{Klessen+05}, where currents moving in opposite directions meet and compress the gas, while {\it reducing} their velocity. This can also be understood as a consequence that, by momentum conservation, {\it dense gas is expected to move more slowly} than diffuse gas, rather than at high velocities. Therefore, high-speed isolated clouds colliding against each other is not likely to be the norm for the formation of high-mass star-forming regions. Rather, large-scale infalling {\it flows} are more physically feasible.

%Collision of filaments: Fukui+15

%Fast HI gas collision: Maeda+21

%Collision of GMCs: Furukawa+09

\subsubsection{The monolithic global collapse model} \label{sec:monolithic_GC}

\citet[] [hereafter KM20] {Krumholz_McKee20} have presented a comparison between simple analytical models representing various scenarios of cloud evolution of star formation. Specifically, KM20 considered models representing a ``static cloud'' (ST), a ``conveyor belt'' (CB) system, an ``increasing star formation efficiency'' (IE) model, and a ``global collapse'' (GC) scenario, the latter said to represent GHC. For the CB and GC models, they considered variants with ``rapid dispersal'' (CBD and GCD, respectively). They then tested whether appropriate parameters can be found for the various models in order to meet various observational constraints on the clouds' star formation activity, such as the clouds' star formation histories, the efficiency over a free-fall time, the galaxy-wide SFR, and all of them combined --- i.e., whether unique sets of parameters for each model can satisfy all constraints at the same time. 

KM20 concluded that only the CB model is able to satisfy all three constraints simultaneously, thanks to that model's property of maintaining a continuous accretion flow onto the dense hubs via filamentary accretion, so that the density of a hub remains nearly stationary, while the density increases rapidly in their GC model, forcing hubs to convert all of their mass into stars in roughly one free-fall time.

However, it must be noted that the GC model of KM20 does {\it not} actually represent the observed behavior of numerical simulations of clouds undergoing GHC, nor the conclusions of articles reporting them, since it represents an {\it isolated and monolithic} collapse, rather than an accreting, hierarchical one. Indeed, simulations such as those of \citet{Colin+13}, GV14, and \citet{GS_VS20} manifestly develop conveyor-belt types of flows, which constitute multi-scale gravitationally-driven flows, precisely such as those represented by KM20's CB model. In contrast, the GC models of KM20 lack the accretion onto each stage of the collapse hierarchy, which is an essential feature of the GHC scenario ever since its inception \citep[e.g.,] [] {VS+09, VS+10}. Thus, {\it the GHC scenario is best represented by KM20's conveyor belt models}, rather than by their own global collapse model, and therefore KM20's own modeling supports the GHC scenario.

\subsection{Evolution and convergence of the TS and GHC models} \label{sec:model_convergence}

The early versions of the TS scenario assumed that molecular clouds in general were globally in near virial equiibrium, supported by turbulent motions, and thus implicitly assumed that the turbulent motions were confined to scales much smaller than the clouds' sizes \citep[e.g.,] [] {Zuck_Evans74}, implying that the only structures able to undergo gravitational contraction were the dense molecular cloud cores \citep[e.g.,] [] {Pad_Nord02}, of size $\sim 0.1$ pc. In more recent times, this view has evolved by recognizing that the scale of the turbulent motions, rather than being small with respect to the clouds, can extend to sizes comparable or even larger than those of the clouds themselves. This has led to proposals such as that by \citet{Evans+21}, that the majority of clouds are not gravitationally bound, and thus perhaps are transient, or that collapsing cloud substructures are formed by large-scale supernova-driven inflows into the clouds \citep[e.g.,] [] {Padoan+16, Ganguly+22}. That is, the clouds are not precisely {\it supported}, but rather are dynamical entities, although not undergoing cloud-scale gravitational contraction. And, within this context, simulations on scales $\sim 5$ pc \citep[e.g.,] [] {Wang+10, Collins+23} have found that the material making up collapsed objects comes from distances up to a few parsecs away, constituting ``clump-fed'' accretion. Therefore, the scale of the collapsing material has increased to up to a few parsecs even within the context of the TS scenario.

On the other hand, GHC has evolved by considering its insertion into the very-large scale Galactic dynamics and the role of continuous external driving from SNe, as discussed in Sec.\ \ref{sec:global} of this work. This continuous {\it external} driving implies that, at least, {\it some} fraction of a cloud's mass may not be participating in the global collapse, and instead it may escape the cloud's body \citep[e.g.,] [] {AV22}. That is, at present, the GHC model explicitly acknowledges that not {\it all} of the material making up a cloud or clump may be participating of the collapse, and that at large scales ($\gtrsim 500$ pc for Milky Way parameters), rotational support must be acknowledged. However, within GHC, the scale of the infall flow can easily extend up to scales of tens of parsecs or more.

Therefore, the two models appear to be converging, with TS acknowledging that structures of scales larger than that of cores (up to $\sim$ a few pc) may be undergoing gravitational contraction, while GHC acknowledging that not the whole mass of a cloud need to be participating in the gravitational collapse flow, 
%{\cyan but rather may be in the process of dispersal after a strong star formation episode}. 
as its turbulent nature and the constant hitting by external driving from nearby supernovae can erode/disperse it, causing some of its material to escape the infall flow.

\section{Summary and conclusions} \label{sec:concls}

In this paper we have presented a detailed comparison between the turbulent support (TS) and global hierarchical collapse (GHC) models for molecular cloud evolution and star formation. This comparison is aimed at dissipating some confusion that has arisen recently regarding the claims and implications of the two models, and at suggesting tests that can help discriminating between the two.

The two models attempt to explain essentially the same set of properties of molecular clouds, their substructures, and their star formation activity, but in terms of two very different interpretations: TS, in its original form, assumes that {\it strongly} supersonic turbulence prevents gravitational contraction of the clouds at all scales larger than the ``core scale'', $\sim 0.1$ pc, while GHC assumes that the cloud dynamics is dominated by gravity (either local or external), and that the disruptive, {\it moderately} supersonic, truly turbulent (i.e., disorganized) motions constitute a minority of the kinetic energy budget of the clouds (although they are still important for the production of nonlinear density fluctuations), and are possibly driven by gravity itself.

Therefore, in this paper, we first summarized the premises of each model (Sec.\ \ref{sec:premises}), and then discussed how various properties of the clouds are interpreted in the context of each (Sec.\ \ref{sec:interpretation}). Next, in Sec.\ \ref{sec:sim_diff} we discussed the similarities and main differences between the two models, to then describe, in Sec.\ \ref{sec:tests}, possible tests to discriminate between them, based on new predictions and implications from each model. 

Our main conclusions are as follows:

\begin{itemize}

\item The TS and GHC models attempt to explain a similar body of observations, under the assumptions of either turbulent support or infall domination, respectively.

\item Both models incorporate turbulence and gravity, but in different proportions and at different scales. The TS scenario assumes that the linewidths of molecular tracers consist exclusively of turbulence, which is {\it strongly} supersonic ($\Ms \sim 10$--30) at the scale of GMCs, and which only becomes subsonic below scales $\sim 0.1$ pc, typical of dense cores. On the other hand, under GHC, the linewidths consist of a combination of infall and turbulent motions, the latter being only {\it moderately} supersonic ($\Ms \sim 3$--10) at the GMC scale.

\item One of the main pending issues to determine which model best describes the dynamics of MCs is the scale at which the motions in molecular structures become dominated by gravity. TS proposes this domination occurs at core ($\sim 0.1$ pc) scales, while GHC proposes it occurs at the entire cloud scale, starting long before the first events of star formation, since the clumps and filaments themselves are the result of gravitational contraction. Moreover, it proposes that collapse is multi-scale, and propagates ``downwards'' in scale, so that smaller, lower-mass objects initiate their local collapse at later times. Finally, GHC assumes a gravitationally-driven multi-scale accretion flow, such that each level of the cloud-clump-filament-core hierarchy accretes from its parent structures.

\item In determining the epoch and scale of gravity domination it is crucial to keep in mind that the central parts of collapsing objects may be locally unbound, as they may just be undergoing compression due to the infall of the larger-scale collapsing material.

\item A stationary description such as that of TS can be appropriate when large enough temporal or spatial scales are considered, so that entire cloud ensembles or many infall/dispersal cycles are included. An estimate of the required spatial averaging scale is given by the size scale at which measurements of the depletion time $\tau_{\rm dep}$ centered on gas (CO) and SF (H$\alpha$) peaks coincide, $\gtrsim 300$ pc.

\item Another important test to discriminate between the models is whether strong shocks are present on one side of well-defined density structures. Such shocks, with a Mach number $\Ms \sim \left(\rho_2/\rho_1\right)^{1/2}$, are necessary to form structures of density $\rho_2$ from a medium initially at density $\rho_1$ under TS in nearly isothermal molecular clouds. No such strong shocks are expected in the GHC scenario.

\item Other models currently under consideration by the community can be accommodated within either one of TS or GHC. In particular, the Turbulent Core and the Inertial Inflow models can be considered as consistent with the TS scenario, while the Hub-Filament System and Cloud-Cloud Collision models are consistent with GHC.

A final conclusion is that, in order to characterize the flow, and to distinguish between the models, it is essential to investigate the flow properties in the {\it surroundings} of the density structures. Features such as the presence of strong shocks and the energetics of the structures are determined by the external conditions, and it is here where the models make different predictions.

\end{itemize}

\section{Acknowledgments}
\label{sec:ack}
We thankfully acknowledge useful and insightful comments from Lee Hartmann, Roberto Galv\'an-Madrid, and Fabian Heitsch, as well as a thoughtful report from an anonymous referee. The simulation surface density plot in Fig. \ref{fig:straight-arm_sim} has been created using the publicly available SPH visualization tool SPLASH by D.J. Price \citep{2007PASA...24..159P}. E.V.-S.\ and A.P. acknowledge support from UNAM-PAPIIT grant IG100223.
G.C.G acknowledges support from UNAM-PAPIIT grant IN110824.
A.P. also acknowledges support from the Sistema Nacional de Investigadores of CONAHCyT. A.P. and J.B.P. acknowledge support from the CONAHCyT grant number 86372 of the `Ciencia de Frontera 2019’ program, entitled `Citlalc\'oatl: A multiscale study at the new frontier of the formation and early evolution of stars and planetary systems’. J.B.P. furthermore acknowledges UNAM-DGAPA-PAPIIT support through grant number {\tt IN-114422}. 

\bigskip
\noindent
{\bf DATA AVAILABILITY}

The numerical simulation data presented in this paper will be shared on reasonable request to the corresponding author.

\bibliographystyle{mnras}
\bibliography{refs}

\appendix

\section{An idealized galactic-flow simulation} \label{app:simulation}

In the following, we present highly idealized simulations to understand the formation of the Nessie filament \citep{2010ApJ...719L.185J, 2014ApJ...797...53G}. The hydrodynamical simulations are performed with the N-Body Smoothed Particle Hydrodynamics (SPH) code GADGET3 \citep{2005MNRAS.364.1105S}.

Our model focuses solely on the interaction between the ISM and the analytical spiral arm potential adapted from \cite{2002ApJS..142..261C}. Using a periodic cube of 4 kpc side length as the simulation volume, we inject a slab of gas with a temperature of 1000 K and a velocity of $15 \kms$ along the $x$-axis into the cube. The slab extends 4 kpc along the $y$-axis and 0.5 kpc above and below the $z$-midplane at $z=2$ kpc. The ``spiral'' potential, which is continuous in space, has been straightened and made time-independent by setting the corresponding variables (y-position, time) to constant values. The central region of the now cylindrical potential is placed at the center of the cube at $x=2$ kpc, $z=2$ kpc, and extends all 4 kpc along the $y$ axis. The potential is given by

\begin{equation}
\begin{aligned}
&\Phi_S = - 4\pi G H\rho_0 {\rm exp} \left(-\frac{r - r_0}{R_s}\right) \times \\
&\sum_{n=1}^{3}\left(\frac{C_n}{K_n D_n}\right){\rm cos}(n\gamma)\left[{\rm sech}\left(\frac{K_n z}{\beta_n}\right)\right]^{\beta_n}
\end{aligned}
\end{equation}
where the components dependent on radius are given by $K_n = \frac{nN}{r {\rm sin}(\alpha)}$, $\beta_n = K_n H(1 + 0.4 K_n H)$, $D_n = \frac{1 + K_nH + 0.3(K_n H)^2}{1 + 0.3 K_n H}$ and the time plus radius dependent component is given by $\gamma = N\left[\theta - \Omega_p t - \frac{{\rm ln}(r/r_0)}{{\rm tan}(\alpha)} \right]$. The variable $N$ determines the number of arms, $\alpha$ the pitch angle, $R_s$ the radial scale length of the drop-off in density amplitude of the arms, $\rho_0$ the mid-plane arm density at fiducial radius $r_0$ and finally $H$ the scale height of the stellar arm perturbation. The values for the constants are given by $C(1) = 8/3\pi$, $C(2) = 1/2$ and $C(3) = 8/15\pi$. The values for the parameters we adapted from \cite{2006MNRAS.371.1663D} are $r_0 = 8$ kpc, $R_s = 7$ kpc, $H = 0.18$ kpc, $\alpha = 15^\circ$, $\Omega_p = 2\times10^{-8}$ rad yr$^{-1}$ and $\rho_0 = 1$ cm$^{-3}$.

The injection rate is set to about 0.09 $\Msun$ yr$^{-1}$, resulting in a constant density of about 0.05 $\pcc$ and a surface density of about 1.5 $\Msun$ pc$^{-2}$ for the slab as long as it remains unperturbed. In our idealized setup, we ensure that there is always unperturbed gas of fixed density and extent flowing across the potential by completely removing gas that leaves the simulation cube on the outflow side at $x=4$ kpc, which would otherwise re-enter the simulation cube from the injection side at $x=0$ kpc. The mass of a single SPH particle is set to about 24 $\Msun$. At the given injection rate, this results in a total number of particles of about $2 \times 10^6$ at the end of the simulation runtime of 530 Myr. The hydro-smoothing uses the normal cubic spline kernel with 60 neighbors and a minimum smoothing length of 0.05 pc, which is also used for the fixed gravitational smoothing length. However, the minimum smoothing length is never reached, since exponentially collapsing fragments reach the density threshold for star formation before that. The Courant factor is set to 0.2.

We follow the time evolution of the slab as it interacts with the spiral potential, focusing on the accumulation of gas within the potential and the subsequent dispersal when densities become high enough for SN feedback to occur due to star formation. %Intermediate states, where much gas has accumulated but feedback is not yet active, are potential candidates for Nessie-like filaments. 
The star formation model is adapted from \cite{2011MNRAS.417.1318D} and is normally used in global simulations of galaxies. We have modified some of the model parameters to account for the smaller spatial scales and better mass resolution in our cube compared to full galaxy simulations. This includes a higher star formation threshold of $3 \times 10^4 \pcc$ compared to the original $10^3 \pcc$ and a smaller region in which particles are checked for collapse (7.5 pc instead of 15 pc). For a region to be considered as collapsing, the central particle density must be above the star formation threshold, the flow inside the region must be convergent and the particles must be gravitationally bound. The total energy from stellar feedback is given by
\begin{equation}
E_{SN} = \frac{\epsilon M_{{\rm H}_2}}{160 M_{\sun}}10^{51}\, {\rm ergs},
\label{eq:sn_energy}
\end{equation}
with each SN contributing $10^{51}$ ergs of energy and with a SN rate of one per 160 $\Msun$ of stars formed. This assumes a Salpeter initial mass function with stellar masses in the range of 0.1 to 100 $\Msun$. The star formation efficiency $\epsilon$ is set to $\epsilon = 10$\%.  Feedback is injected as approximately 1/3 thermal and 2/3 kinetic energy into the gas remaining after $\epsilon$ percent of the gas in the collapsing region has been converted into a single star cluster sink particle. The feedback is instantaneous, no delay is added between sink formation and SN onset. There is no additional fixed rate SN injection at random positions (Type Ia SNe), so feedback always originates from the sink particles (Type II SNe). Since our simulation does not include the formation of molecular hydrogen, we take the total gas mass of a collapsing region for $M_{H_2}$. The cooling function we use is a parameterization adapted from \citet{VS+07}.

\section{Simulations of decaying turbulence in the warm neutral medium}
\label{app:phantom_simulation}

Here we summarize the main features of the two simulations presented in Fig.\ \ref{fig:sigma_N_maps} on the formation and evolution of GMC's and filaments in the WNM. We use the smoothed-particle hydrodynamics (SPH) code \textsc{PHANTOM} \citep{Price+2018} to perform two simulations of decaying turbulence with and without gravity. For these simulations we used the default phantom-setup \textit{cluster}, based on the work of \cite{Bate_Bonnell_Bromm2003}, combined with a close-packed arrangement of $128 \times 148 \times 157 \approx 3 \times 10^6$ particles in a box of $256$ pc per side. Periodic boundary conditions are used for the hydrodynamics, but self-gravity is solved with isolated boundary conditions. The initial density and temperature were set at $n(t=0) = 3\, \pcc$ and $T(t=0) = 730$ K, respectively, the latter being the thermal equilibrium temperature at the initial density. 

Stellar (sink) particles are formed once the particle density reaches $4.7 \times 10^{5}\, \pcc$. The cooling and heating processes are included via fitted functions from \cite{Koy_Inu02}, with the typographical correction given by \citet{VS+07}. We kept the default pure-solenoidal forcing files of the \textit{cluster} setup, which was generated on a $64^{3}$ uniform grid in Fourier space with a power spectrum of $P(k)\propto k^{-4}$, but corrected by a normalization factor which produces a velocity dispersion of $\sigma \approx 5 \kms$ at time $t=0$. The time studied in both cases corresponds to $t=14.7$ Myr. These simulations do not include any form of stellar feedback.

\end{document}